\documentclass[modern]{aastex63}
\newcommand{\re}{\textcolor[rgb]{0,0,0}}
\newcommand{\ree}{\textcolor[rgb]{0,0,0}}
\newcommand{\reee}{\textcolor[rgb]{0,0,0}}

\def\gtsima{$\;\buildrel > \over \sim \;$}
\def\simgt{\lower.5ex \hbox{\gtsima}}
\def\ltsima{$\;\buildrel < \over \sim \;$}
\def\simlt{\lower.5ex \hbox{\ltsima}}
\def \CHp{\ifmmode{\rm CH^+}\else{$\rm CH^+$}\fi}
\def \HH{\ifmmode{\rm H_2}\else{$\rm H_2$}\fi}
\def \Cp{\ifmmode{\rm C^+}\else{$\rm C^+$}\fi} 
\def \cc    {\ifmmode{\,{\rm cm}^{-3}}\else{$\,{\rm cm}^{-3}$}\fi}
\def \dens{\ifmmode{n_{\rm H}}\else{$n_{\rm H}$}\fi}
\def \kms   {\ifmmode{\,{\rm km}\,{\rm s}^{-1}}\else{km s$^{-1}$}\fi} 

\journalinfo{Accepted for publication in ApJ, 2024 Aug 12}

\begin{document}

\title{The densities in diffuse and translucent molecular clouds: estimates from
observations of C$_2$ and from 3-dimensional extinction maps}

\author{David A. Neufeld}
\affiliation{William H.\ Miller Department of Physics \& Astronomy, Johns Hopkins University, Baltimore, MD 21218, USA}
\author{Daniel E. Welty}
\affiliation{Space Telescope Science Institute, Baltimore, MD 21218, USA}
\author{Alexei V. Ivlev}
\affiliation{Max-Planck Institute for Extraterrestrial Physics, 85748 Garching, DE}
\author{Paola Caselli}
\affiliation{Max-Planck Institute for Extraterrestrial Physics, 85748 Garching, DE}
\author{Gordian Edenhofer}
\affiliation{Max Planck Institute for Astrophysics, 85748 Garching, DE}
\author{Nick Indriolo}
\affiliation{AURA for ESA, Space Telescope Science Institute, Baltimore, MD 21218, USA}
\author{Marta Obolentseva}
\affiliation{Max-Planck Institute for Extraterrestrial Physics, 85748 Garching, DE}
\author{Kedron Silsbee}
\affiliation{University of Texas, El Paso, TX 79968, USA}
\author{Paule Sonnentrucker}
\affiliation{European Space Agency, ESA office at STScI, Baltimore, MD 21218, USA}
\author{Mark G. Wolfire}
\affiliation{University of Maryland, College Park, MD 20742, USA}
\correspondingauthor{David A. Neufeld}

\begin{abstract}

Newly-computed collisional rate coefficients for the excitation of C$_2$ in collisions with H$_2$, 
presented recently by Najar and Kalugina (2020), are significantly larger than the values 
adopted previously in models for the excitation of the C$_2$ molecule, a widely used
probe of the interstellar gas density.
With these new rate coefficients, 
we have modeled the C$_2$ rotational distributions inferred from visible and ultraviolet
absorption observations of electronic transitions of C$_2$ towards a collection of
46 nearby background sources.  The inferred gas densities in the foreground interstellar clouds 
responsible for the observed C$_2$ absorption are a factor 4 to 7 
smaller than those inferred previously, a direct reflection of the larger collisional 
rate coefficients computed by Najar and Kalugina (2020).  These lower density estimates are
generally in good agreement with the peak densities inferred from 3D extinction maps
for the relevant sightlines.  In cases where H$_3^+$ absorption has also been observed and 
used to estimate the cosmic-ray ionization rate (CRIR), our estimates of the
latter will also decrease accordingly because the H$_3^+$ abundance is a function of the
ratio of the CRIR to the gas density.   

\end{abstract}


\section{Introduction}

The diatomic carbon molecule, C$_2$, has been widely used as a density estimator in the
diffuse and translucent interstellar clouds.  As discussed by 
Chaffee et al.\ (1980) and van Dishoeck \& Black (1982; hereafter vDB82), 
the rotational distribution of C$_2$ in its ground vibrational and electronic state 
-- which can be measured by means of absorption line observations of 
several electronic bands in the visible and ultraviolet spectral regions -- reflects a
competition between inelastic collisions with He or H$_2$ and pumping by the
interstellar radiation field (ISRF), primarily in the Phillips A -- X band 
near 1~$\mu$m.  Figure 1 shows an example rotational diagram obtained toward the
star HD 24534 (Sonnentrucker et al.\ 2007; hereafter S07), in which log$_{10}(f_J/g_J)$ 
is plotted against $E_J/k$, where $f_J$ is the 
fractional population in the state of rotational quantum number, $J$, and $E_J$ and
$g_J = 2J+1$ are the energy and degeneracy of that state.  For C$_2$, only even-$J$ states
are present because the wavefunction must be symmetric under exchange of the two identical
(bosonic) nuclei.   
\begin{figure}[t!]
\includegraphics[angle=0,width=6 true in]{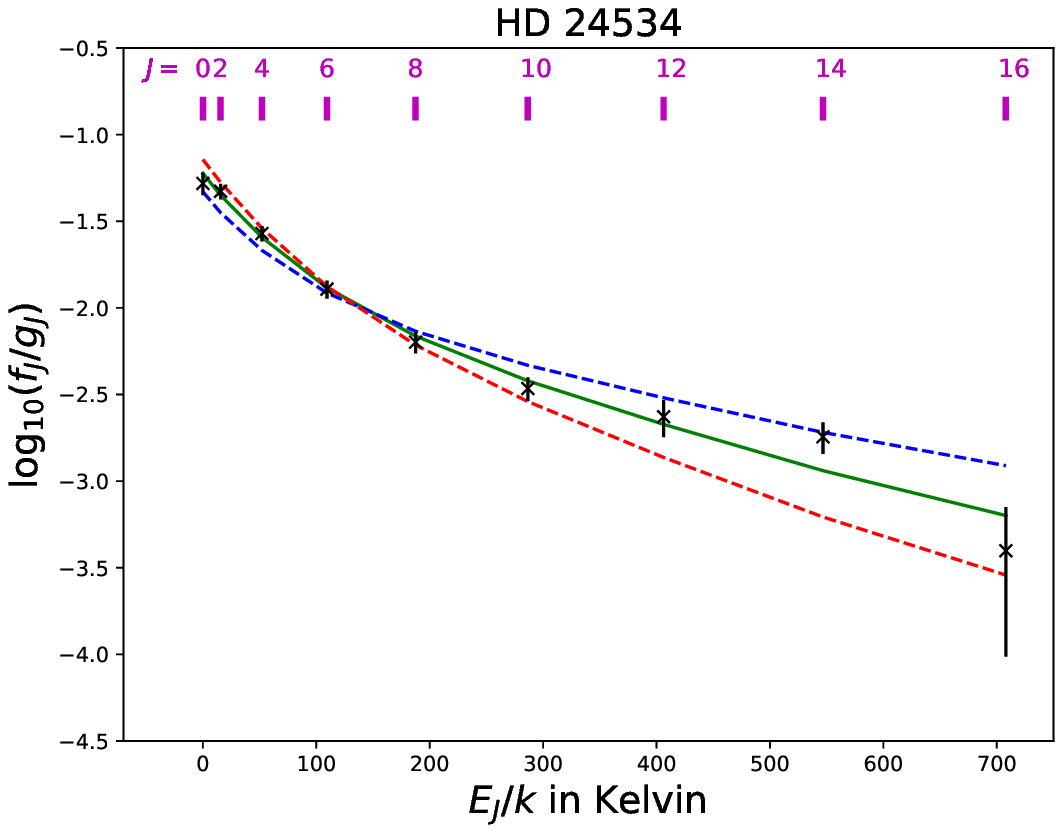}
\figcaption{Rotational diagram from C$_2$ absorption line observations (S07)
toward HD 24534.  Green curve: best fit, obtained for
$T = 48$~K and $n_{\rm H} = 70 \, \rm cm^{-3}$, using the excitation
model described in Section 3.  Red dashed curve: fit for $T = 48$~K 
and $n_{\rm H} = 98 \, \rm cm^{-3}$.  Blue dashed curve: fit for $T = 48$~K 
and $n_{\rm H} = 50 \, \rm cm^{-3}$.}
\end{figure}
For low $J$, the populations
are controlled by collisional excitation, and the (negative) slope of the rotational diagram is 
inversely proportional to the kinetic temperature of the gas.  For high $J$, the effects
of radiative pumping become important, and the slope of the curve becomes less negative.
The value of $E_J$ above which the slope becomes less negative is accordingly an increasing 
function of density.  The green curve is a best fit to the observed data, obtained using
the methods described in Section 3 below for 
a gas kinetic temperature, $T = 48$~K, and a gas density, $n_{\rm H} = 70 \, \rm cm^{-3}$,
defined here as the volume density of hydrogen nuclei, 
$n({\rm H}) + n({\rm H}^+) + 2n({\rm H}_2)$.  Red and blue curves show respectively 
the rotational diagrams expected for assumed densities that are 40\% larger and 40\% smaller
than the best-fit values.

Following Chaffee et al.\ (1980), and in the absence of any quantum-mechanical 
calculations of relevance, vDB82 adopted a very simple approximation for the excitation rates 
for C$_2$.  They assumed that the colllisional cross-section, $\sigma(2,0)$, for 
de-excitation from $J=2$ to $J=0$ was $2 \times 10^{-16}\,\rm cm^{2}$ independent of energy.  
For other collisionally-induced deexcitations from $J_u$ to $J_l$, they adopted a cross-section
$\sigma(J_u,J_l)=5\sigma(2,0)/(2J_u+1)$ for $J_u=J_l+2$ and $\sigma(J_u,J_l)=0$ otherwise.
They also computed the radiative excitation matrix, $Y(J_i,J_f)$, which describes the probability
that radiative pumping in an initial rotational state $J_i$ will be followed by a 
radiative cascade that ultimately leads to a reentry into the ground electronic and vibrational state
in rotational state $J_f$.  Together with an estimate of the pumping rate, $W_J$, (dependent
on the ISRF and the relevant oscillator strengths) and the inclusion of spontaneous radiative 
decay within the ground vibrational state (generally negligible because the pure rotational
transitions are dipole-forbidden), this allowed vDB82 to compute the fractional level 
populations of C$_2$,
$f_J$, expected in equilibrium, as a function of density and temperature.

Over the subsequent four decades, the 
treatment of C$_2$ excitation described by
vDB82 has been widely used to determine densities in the interstellar medium and has been 
implemented in an online tool.  \ree{Because
vDB82 adopted the same collisional excitation rates for excitation by all collision partners
(e.g. H, H$_2$ and He), the densities thereby determined were considered to be the total
particle densities, $n = n({\rm H}) + n({\rm H_2}) + n({\rm He})$.}
Casu and Cecchi-Pestillini (2012, hereafter CC12) presented the results from 
a revised excitation model in which the radiative excitation matrix and pumping rate
were recomputed to reflect recent improvements in the molecular data and the 
collisional rate coefficients were updated beyond the simple  estimates 
introduced by  Chaffee et al.\ (1980).  Here, they made use of close-coupling calculations 
of the rate coefficients for excitation of C$_2$ by He (Najar et al.\ 2008) and para-H$_2$ 
(Najar et al.\ 2009).    
As discussed in Section 2,
a recent high-accuracy calculation of the rate coefficients
for excitation of C$_2$ by both ortho- and para-H$_2$ (Najar \& Kalugina 2020, hereafter NK20)
allows the use of C$_2$ as a density estimator to be placed on a sounder footing and
leads to a significant decrease in the inferred gas densities. 

With the advent of high-quality observational data obtained with {\it Gaia}, an alternative 
method of density estimation has emerged.  Distance and G-band extinction measurements have been
obtained toward a large sample of stars in the Galaxy, allowing 3-dimensional maps of the extinction density
(i.e. extinction per unit path length) to be constructed (e.g.  
Lallement et al.\ 2019; Leike et al.\ 2020, hereafter L20; Edenhofer et al.\ 2023, hereafter E23).  
For a given assumed 
gas-to-extinction ratio, these maps may be used to determine the gas density, n$_{\rm H}$,
on a 3-dimensional grid of Galactic positions $(X, Y, Z)$.  
In Section 3, we discuss the comparison between the density 
estimates obtained from the 3D extinction maps with those obtained using our 
updated analysis of C$_2$ absorption spectra: results are given for \re{23} sightlines
toward background sources located within 1.25 kpc of the Sun.    
In section 4, we discuss briefly the implications of our lower density estimates 
in translucent interstellar clouds for the cosmic-ray ionization rates (CRIR) inferred from 
observations of H$_3^+$.  A more sophisticated analysis, performed with the use of models
that account for the observed 3D distribution of gas and hot stars, is presented in a companion 
paper by Obolentseva et al.\ (2024).

\section{$\bf C_2$ density estimates}

Our treatment of the excitation of C$_2$ follows closely the methodology of vDB82.  We solve 
the equations of statistical equilibrium (vDB82, equation 27) as a function of gas kinetic 
temperature and density, but we use the collisional rate coefficients computed by NK20
in place of the simple formula introduced by Chaffee et al.\ (1980).  Figure 2 shows a
comparison of the rate coefficients for collisional deexcitation to the ground state
($J=0$) from the first excited 
rotational state ($J=2$).
  
The black curve shows the value adopted by vDB82. The red and blue curves 
show the NK20 results for collisions with para-H$_2$ and 
ortho-H$_2$ respectively, and the magenta curve shows the earlier results
obtained for He (Najar et al.\ 2008).  
As noted by NK20, the
rate coefficient for excitation by ortho-H$_2$ is much larger than that for para-H$_2$, 
a behavior that is observed in many other systems.  Thus, the availability of the new
NK20 rate coefficients with their first high-quality calculation for excitation by 
ortho-H$_2$ has a significant impact on our understanding of C$_2$ excitation.
The cyan curve shows the weighted average appropriate to an H$_2$ ortho-to-para ratio in equilibrium
at the gas kinetic temperature.  The green curve includes the additional
effect due to He for an assumed 
He abundance of 0.2 relative to H$_2$, and is defined such that the total deexcitation rate
per C$_2$ molecule, including the effects of helium, is $q(2,0)n({\rm H_2}$).  

\begin{figure}[t!]
\includegraphics[angle=0,width=\textwidth]{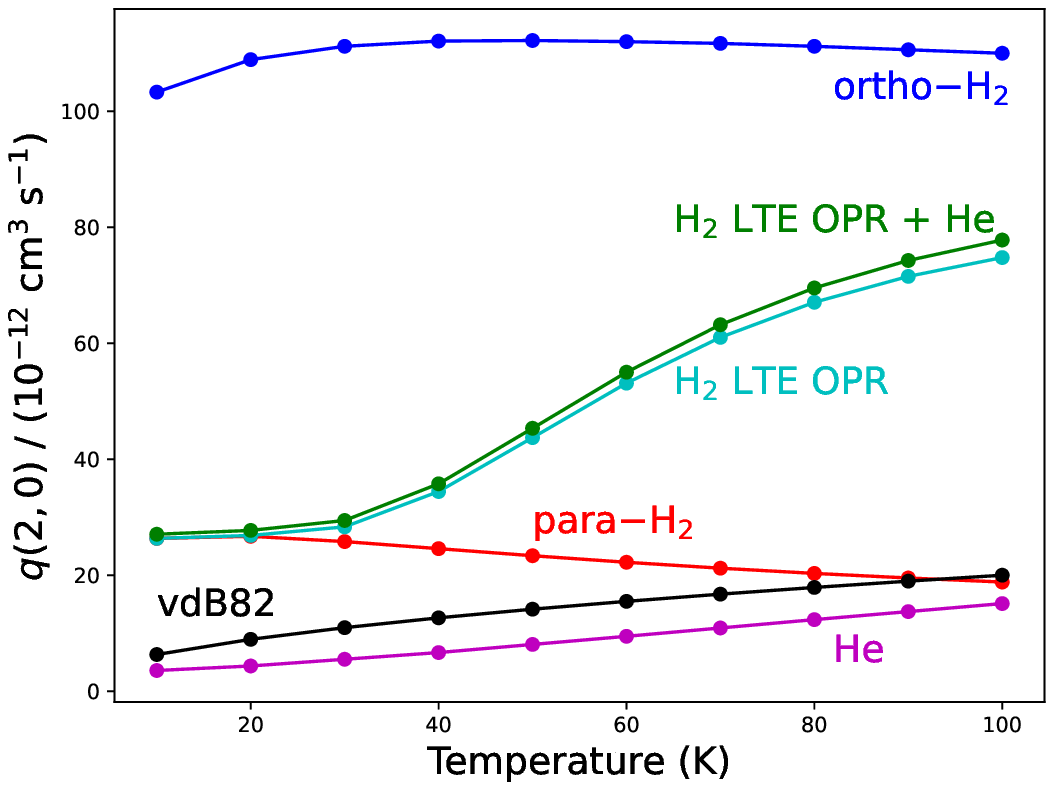}
\figcaption{Comparison of rate coefficients for the collisional de-excitation
of C$_2$ $J = 2$ to $J = 0$ computed for various collision partners: 
para-H$_2$ (from NK20 in red), ortho-H$_2$ (from NK20 in blue), 
and He (from Najar et al.\ 2008 in magenta).  Cyan points: weighted average of NK20 results
for an H$_2$ ortho-to-para ratio in LTE.  Green points: weighted averaged
per H$_2$ molecule, including the additional contribution of He for a He/H$_2$
abundance ratio of 0.2.  Black points: simple
expression adopted by vDB82.  }
\end{figure}
Moreover, while the cross-sections $\sigma(J_u,J_u-2)$
adopted by vDB82 are a monotonically decreasing function of $J_u$, the values of 
$\sigma(4,2)$ computed by NK20 are larger than those for $\sigma(2,0)$, a behavior noted
in previous studies (e.g. Phillips 1994).   Figure 3 shows
the ratio $q(J_u,J_u-2)/q(2,0)$ computed by NK20 at a temperature of 60 K, 
both for collisions with para-H$_2$ (red) and ortho-H$_2$ (blue).
For comparison, the dependence assumed by vDB82 
is shown by the black curve.
\begin{figure}[t!]
\includegraphics[angle=0,width=\textwidth]{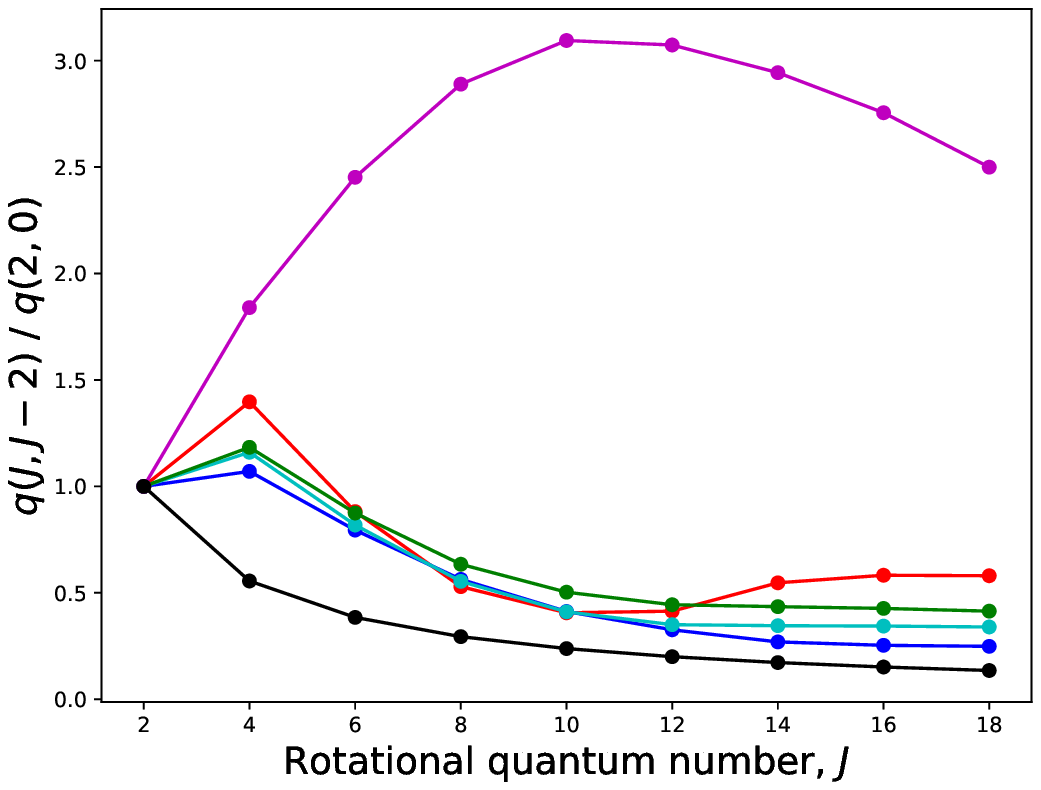}
\figcaption{The ratio $q(J_u,J_u-2)/q(2,0)$ computed by NK20 at a temperature of 60 K, 
both for collisions with para-H$_2$ (red), ortho-H$_2$ (blue), and He (magenta).
Cyan points: weighted average of NK20 results
for an H$_2$ ortho-to-para ratio in LTE.  Green points: weighted averaged
per H$_2$ molecule, including the additional contribution of He for a He/H$_2$
abundance ratio of 0.2.  Black points: dependence assumed by vDB82.}
\end{figure}

We have also adopted the revised radiative excitation matrix, $Y(J_i,J_f)$, 
recomputed by CC12 for the ISRF assumed by vDB82 and given by:

$${4 \pi J_\nu \over h\nu} = \re{1.54 \times 10^{-5}\, 
\lambda_{\mu{\rm m}}^{2.7}\,\rm} {\rm photons\, cm^{-2} \, s^{-1} \, Hz^{-1}},\eqno(1)$$
where $J_\nu$ is the mean (angle averaged) intensity at frequency $\nu$,
and $\lambda_{\mu{\rm m}}$ is the wavelength in micron.

\re{Over the 0.77 -- 1.21$\mu$m range most relevant for C$_2$ excitation, 
the $\lambda_{\mu{\rm m}}^{2.7}$ wavelength dependence adopted by vDB82 is in excellent
agreement with a subsequent study of the ISRF by Mathis et al.\ (1983).  However, the 
absolute value of the ISRF adopted by vDB82 
is a factor\footnote{In determining
this factor, we made use of the zodi-subtracted DIRBE all-sky map, 
available in an equal area pixelation (HEALPix) at 
\newline
https://irsa.ipac.caltech.edu/data/Planck/release\_1/external-data/external\_maps.html,
\newline
and the DIRBE Band 1 spectral response function at 
\newline
https://lambda.gsfc.nasa.gov/product/cobe/dirbe\_ancil\_sr\_get.html} 
1.39 smaller than that measured (0.496 MJy sr$^{-1}$)
by the DIRBE instrument on {\it COBE} at 1.25 $\mu$m (Hauser et al.\ 1998). 
Accordingly, we
have scaled the pumping rate given by CC12 ($W_J=3.1 \times 10^{-9}\,\rm s^{-1}$, 
independent of $J$) by
a factor 1.39 to obtain the revised estimate $W_J= 4.3 \times 10^{-9}\,\rm s^{-1}$.}

We have used our C$_2$ excitation model to fit the fractional C$_2$ populations, $f_J$,
obtained from absorption line spectroscopy performed toward 46 background sources for which C$_2$
has been detected in rotational states extending up to $J = 12$ or higher: estimates
of $f_J$ with $J \ge 12$ are needed to provide enough leverage on the rotational
diagram to constrain the gas density.  Our sample -- listed in Table 1 --
was assembled from a combination of previously-published results and fits 
to archival data.  The latter were downloaded from the relevant archive and their spectra
normalized to the continuum.  We then performed profile fits to the normalized spectra, 
using the wavelengths and oscillator strengths adopted by S07.  The Doppler parameter, $b$,
was assumed to be the same for all transitions, with a value of 
$1 \, \rm km \, s^{-1}$ being typically adopted.  The derived column densities show little 
dependence on the value of $b$, because the lines 
are typically optically-thin.
In most cases, all the lines were fitted simultaneously with a single component. 
In cases where C$_2$ was not detected in rotational states as high as $J = 12$, the model 
typically yields little useful information:  only lower limits on the density are
obtained, and these typically lie below the values obtained for other sources 
where states of $J \ge 12$ 
are available.

\begin{deluxetable}{lccccccccccllc}
\tablenum{1}
\tablewidth{9.8 true in}
\tabletypesize{\scriptsize}
\rotate
\tablecaption{C$_2$ column densities ($10^{13}\,\rm cm^{-2}$), and derived temperatures (K) 
and densities (cm$^{-3}$)}
\tablehead{\cr
Source & N(0) & N(2) & N(4) & N(6) & N(8)
    & N(10) & N(12) & N(14) & N(16) & N(18) & $n_{\rm H}({\rm C}_2)$ & $T({\rm C}_2)$ & Ref$^{a}$}
\startdata
Cernis 52& 0.88 $\pm$ 0.13& 2.79 $\pm$ 0.30& 2.54 $\pm$ 0.32& 1.54 $\pm$ 0.22& 0.98 $\pm$ 0.38& 0.66 $\pm$ 0.29& 0.63 $\pm$ 0.21& ...& ...& ...& $94_{-6}^{+6}$ & $32_{-5}^{+5}$ & IG11 \cr
HD 24534& 0.16 $\pm$ 0.02& 0.72 $\pm$ 0.06& 0.74 $\pm$ 0.06& 0.51 $\pm$ 0.05& 0.33 $\pm$ 0.04& 0.22 $\pm$ 0.03& 0.18 $\pm$ 0.04& 0.16 $\pm$ 0.03& 0.04 $\pm$ 0.03& ...& $70_{-6}^{+4}$ & $46_{-7}^{+9}$ & S07 \cr
HD 27778& 0.15 $\pm$ 0.02& 0.59 $\pm$ 0.05& 0.62 $\pm$ 0.05& 0.52 $\pm$ 0.04& 0.34 $\pm$ 0.04& 0.18 $\pm$ 0.03& 0.15 $\pm$ 0.02& 0.11 $\pm$ 0.02& 0.09 $\pm$ 0.02& 0.06 $\pm$ 0.02& $64_{-4}^{+2}$ & $50_{-6}^{+8}$ & S07 \cr
HD 29647& 2.08 $\pm$ 0.04& 3.68 $\pm$ 0.05& 1.79 $\pm$ 0.05& 0.82 $\pm$ 0.04& 0.65 $\pm$ 0.05& 0.63 $\pm$ 0.11& 0.54 $\pm$ 0.05& ...& ...& ...& $122_{-6}^{+8}$ & $10_{-1}^{+2}$ & ARC \cr
HD 34078& 0.24 $\pm$ 0.02& 1.05 $\pm$ 0.04& 1.23 $\pm$ 0.04& 1.22 $\pm$ 0.04& 0.98 $\pm$ 0.04& 0.75 $\pm$ 0.04& 0.60 $\pm$ 0.04& ...& 0.22 $\pm$ 0.05& ...& $56_{-10}^{+28}$ & $97_{-19}^{+26}$ & ARC \cr
NGC2024 1& 1.22 $\pm$ 0.07& 4.09 $\pm$ 0.12& 4.05 $\pm$ 0.13& 2.43 $\pm$ 0.12& 1.58 $\pm$ 0.12& 1.08 $\pm$ 0.11& 0.43 $\pm$ 0.12& 0.50 $\pm$ 0.12& ...& ...& $96_{-4}^{+6}$ & $41_{-4}^{+4}$ & ARC \cr
HD 46202& 0.10 $\pm$ 0.02& 0.47 $\pm$ 0.04& 0.43 $\pm$ 0.04& 0.37 $\pm$ 0.03& 0.28 $\pm$ 0.03& 0.14 $\pm$ 0.03& 0.11 $\pm$ 0.05& ...& ...& ...& $80_{-12}^{+36}$ & $64_{-14}^{+17}$ & STIS \cr
HD 46223& 0.09 $\pm$ 0.02& 0.34 $\pm$ 0.03& 0.38 $\pm$ 0.03& 0.35 $\pm$ 0.03& 0.19 $\pm$ 0.03& 0.09 $\pm$ 0.03& 0.07 $\pm$ 0.03& ...& ...& ...& $164_{-62}^{+773}$ & $84_{-13}^{+17}$ & STIS \cr
Walker 67& 0.43 $\pm$ 0.01& 1.50 $\pm$ 0.01& 1.26 $\pm$ 0.01& 0.79 $\pm$ 0.01& 0.45 $\pm$ 0.01& 0.27 $\pm$ 0.01& 0.21 $\pm$ 0.01& 0.21 $\pm$ 0.01& ...& ...& $102_{-4}^{+6}$ & $33_{-3}^{+4}$ & UVG \cr
HD 62542& 0.69 $\pm$ 0.05& 2.58 $\pm$ 0.14& 2.57 $\pm$ 0.14& 1.63 $\pm$ 0.08& 0.85 $\pm$ 0.04& 0.48 $\pm$ 0.03& 0.29 $\pm$ 0.03& 0.20 $\pm$ 0.02& 0.10 $\pm$ 0.02& 0.10 $\pm$ 0.02& $104_{-4}^{+6}$ & $41_{-5}^{+5}$ & W20 \cr
HD 63804& 0.48 $\pm$ 0.02& 1.91 $\pm$ 0.03& 1.73 $\pm$ 0.07& 1.27 $\pm$ 0.03& 0.88 $\pm$ 0.03& 0.50 $\pm$ 0.03& 0.38 $\pm$ 0.03& 0.33 $\pm$ 0.06& ...& ...& $82_{-2}^{+4}$ & $42_{-4}^{+5}$ & UVG \cr
HD 73882& 0.71 $\pm$ 0.06& 1.56 $\pm$ 0.09& 0.81 $\pm$ 0.05& 0.46 $\pm$ 0.04& 0.21 $\pm$ 0.03& 0.13 $\pm$ 0.03& 0.12 $\pm$ 0.03& 0.09 $\pm$ 0.03& 0.05 $\pm$ 0.03& ...& $132_{-8}^{+10}$ & $16_{-3}^{+3}$ & STIS \cr
HD 80077& 0.96 $\pm$ 0.21& 3.23 $\pm$ 0.46& 1.77 $\pm$ 0.54& 1.23 $\pm$ 0.49& 0.84 $\pm$ 0.42& 0.94 $\pm$ 0.50& 0.73 $\pm$ 0.42& ...& ...& ...& $88_{-8}^{+12}$ & $20_{-4}^{+5}$ & G99 \cr
HD 112272& 0.06 $\pm$ 0.02& 0.23 $\pm$ 0.02& 0.17 $\pm$ 0.02& 0.27 $\pm$ 0.02& 0.24 $\pm$ 0.02& 0.12 $\pm$ 0.02& 0.21 $\pm$ 0.02& ...& ...& ...& $78_{-50}^{+923}$ & $156_{-138}^{+48}$ & UVP \cr
HD 136239& 0.28 $\pm$ 0.04& 0.93 $\pm$ 0.08& 0.73 $\pm$ 0.13& 0.57 $\pm$ 0.12& 0.25 $\pm$ 0.12& 0.17 $\pm$ 0.06& 0.13 $\pm$ 0.07& 0.10 $\pm$ 0.09& ...& ...& $104_{-6}^{+6}$ & $33_{-4}^{+5}$ & K10 \cr
HD 147084& 0.19 $\pm$ 0.06& 0.76 $\pm$ 0.15& 0.70 $\pm$ 0.23& 0.58 $\pm$ 0.15& 0.57 $\pm$ 0.13& 0.24 $\pm$ 0.14& 0.23 $\pm$ 0.14& 0.17 $\pm$ 0.13& ...& ...& $62_{-8}^{+8}$ & $50_{-14}^{+20}$ & vDB \cr
HD 147888& 0.10 $\pm$ 0.02& 0.33 $\pm$ 0.03& 0.36 $\pm$ 0.03& 0.30 $\pm$ 0.03& 0.18 $\pm$ 0.03& 0.14 $\pm$ 0.02& 0.08 $\pm$ 0.02& 0.05 $\pm$ 0.02& ...& ...& $68_{-2}^{+4}$ & $56_{-7}^{+9}$ & S07 \cr
HD 147889& 0.86 $\pm$ 0.03& 2.71 $\pm$ 0.04& 2.85 $\pm$ 0.05& 2.29 $\pm$ 0.04& 1.59 $\pm$ 0.05& 0.93 $\pm$ 0.07& 0.69 $\pm$ 0.04& 0.31 $\pm$ 0.04& 0.14 $\pm$ 0.03& ...& $72_{-2}^{+4}$ & $54_{-6}^{+7}$ & ARC \cr
HD 147933& 0.12 $\pm$ 0.01& 0.52 $\pm$ 0.02& 0.52 $\pm$ 0.02& 0.30 $\pm$ 0.02& 0.20 $\pm$ 0.02& 0.15 $\pm$ 0.02& 0.08 $\pm$ 0.02& 0.05 $\pm$ 0.02& 0.05 $\pm$ 0.02& 0.04 $\pm$ 0.02& $84_{-4}^{+6}$ & $43_{-5}^{+5}$ & STIS \cr
HD 148184& 0.18 $\pm$ 0.02& 0.70 $\pm$ 0.03& 0.81 $\pm$ 0.04& 0.63 $\pm$ 0.04& 0.39 $\pm$ 0.04& 0.34 $\pm$ 0.03& 0.14 $\pm$ 0.07& ...& ...& ...& $78_{-12}^{+30}$ & $65_{-14}^{+16}$ & UVP \cr
HD 148379& 0.07 $\pm$ 0.04& 0.25 $\pm$ 0.07& 0.24 $\pm$ 0.09& 0.21 $\pm$ 0.09& 0.15 $\pm$ 0.08& 0.11 $\pm$ 0.10& 0.06 $\pm$ 0.05& ...& ...& ...& $70_{-6}^{+4}$ & $48_{-9}^{+12}$ & K10 \cr
HD 149757& 0.08 $\pm$ 0.01& 0.26 $\pm$ 0.02& 0.32 $\pm$ 0.03& 0.25 $\pm$ 0.03& 0.18 $\pm$ 0.03& 0.14 $\pm$ 0.02& 0.11 $\pm$ 0.02& 0.08 $\pm$ 0.02& 0.06 $\pm$ 0.02& ...& $52_{-2}^{+2}$ & $48_{-6}^{+6}$ & S07 \cr
HD 151932& 0.20 $\pm$ 0.04& 0.54 $\pm$ 0.09& 0.56 $\pm$ 0.08& 0.48 $\pm$ 0.10& 0.20 $\pm$ 0.10& 0.24 $\pm$ 0.09& 0.16 $\pm$ 0.14& 0.10 $\pm$ 0.08& ...& ...& $76_{-8}^{+6}$ & $37_{-9}^{+11}$ & K10 \cr
HD 152003& 0.08 $\pm$ 0.02& 0.43 $\pm$ 0.03& 0.47 $\pm$ 0.03& 0.42 $\pm$ 0.03& 0.30 $\pm$ 0.03& 0.26 $\pm$ 0.03& 0.23 $\pm$ 0.07& ...& ...& ...& $58_{-8}^{+22}$ & $74_{-17}^{+23}$ & UVP \cr
HD 152235& 0.12 $\pm$ 0.02& 0.41 $\pm$ 0.04& 0.31 $\pm$ 0.03& 0.35 $\pm$ 0.03& 0.23 $\pm$ 0.03& 0.15 $\pm$ 0.03& 0.17 $\pm$ 0.03& ...& ...& ...& $64_{-14}^{+10}$ & $35_{-13}^{+28}$ & UVP \cr
HD 152236& 0.13 $\pm$ 0.03& 0.36 $\pm$ 0.06& 0.30 $\pm$ 0.06& 0.29 $\pm$ 0.06& 0.13 $\pm$ 0.06& 0.12 $\pm$ 0.10& 0.08 $\pm$ 0.07& ...& ...& ...& $86_{-10}^{+14}$ & $32_{-9}^{+13}$ & K10 \cr

\enddata
\end{deluxetable}

\begin{deluxetable}{lccccccccccllc}
\tablenum{1 (contd.)}
\tablewidth{9.8 true in}
\tabletypesize{\scriptsize}
\rotate
\tablecaption{C$_2$ column densities ($10^{13}\,\rm cm^{-2}$), and derived temperatures (K) 
and densities (cm$^{-3}$)}
\tablehead{\cr
Source & N(0) & N(2) & N(4) & N(6) & N(8)
    & N(10) & N(12) & N(14) & N(16) & N(18) & $n_{\rm H}({\rm C}_2)$ & $T({\rm C}_2)$ & Ref$^{a}$}
\startdata
HD 152270& 0.10 $\pm$ 0.02& 0.28 $\pm$ 0.03& 0.39 $\pm$ 0.03& 0.29 $\pm$ 0.04& 0.13 $\pm$ 0.04& 0.16 $\pm$ 0.04& 0.11 $\pm$ 0.08& ...& ...& ...& $82_{-18}^{+138}$ & $68_{-19}^{+28}$ & UVP \cr
HD 152590& 0.07 $\pm$ 0.01& 0.24 $\pm$ 0.02& 0.25 $\pm$ 0.02& 0.17 $\pm$ 0.02& 0.10 $\pm$ 0.02& 0.08 $\pm$ 0.02& 0.05 $\pm$ 0.02& ...& ...& ...& $88_{-6}^{+6}$ & $46_{-5}^{+6}$ & STIS \cr
HD 154368& 0.41 $\pm$ 0.02& 1.33 $\pm$ 0.05& 1.16 $\pm$ 0.06& 0.83 $\pm$ 0.06& 0.64 $\pm$ 0.06& 0.32 $\pm$ 0.06& 0.18 $\pm$ 0.05& 0.29 $\pm$ 0.08& 0.27 $\pm$ 0.08& ...& $80_{-4}^{+4}$ & $33_{-5}^{+5}$ & K10 \cr
HD 163800& 0.19 $\pm$ 0.03& 0.49 $\pm$ 0.06& 0.64 $\pm$ 0.09& 0.47 $\pm$ 0.09& 0.27 $\pm$ 0.07& 0.19 $\pm$ 0.08& ...& 0.13 $\pm$ 0.09& ...& ...& $78_{-12}^{+46}$ & $52_{-17}^{+22}$ & K09 \cr
HD 169454& 0.95 $\pm$ 0.04& 2.38 $\pm$ 0.06& 1.48 $\pm$ 0.05& 0.68 $\pm$ 0.05& 0.40 $\pm$ 0.05& 0.18 $\pm$ 0.05& 0.18 $\pm$ 0.05& 0.12 $\pm$ 0.08& 0.12 $\pm$ 0.08& ...& $136_{-8}^{+8}$ & $21_{-2}^{+2}$ & K09 \cr
HD 170740& 0.24 $\pm$ 0.10& 0.41 $\pm$ 0.06& 0.42 $\pm$ 0.06& 0.22 $\pm$ 0.07& 0.32 $\pm$ 0.10& 0.13 $\pm$ 0.05& 0.14 $\pm$ 0.05& ...& ...& ...& $70_{-8}^{+10}$ & $31_{-12}^{+15}$ & K10 \cr
HD 172028& 1.00 $\pm$ 0.04& 2.97 $\pm$ 0.06& 2.58 $\pm$ 0.07& 1.86 $\pm$ 0.06& 1.14 $\pm$ 0.06& 0.64 $\pm$ 0.13& 0.32 $\pm$ 0.06& 0.50 $\pm$ 0.07& ...& ...& $94_{-6}^{+6}$ & $35_{-5}^{+5}$ & ARC \cr
HD 179406& 0.24 $\pm$ 0.02& 1.03 $\pm$ 0.03& 1.17 $\pm$ 0.04& 0.88 $\pm$ 0.04& 0.37 $\pm$ 0.03& 0.20 $\pm$ 0.07& 0.12 $\pm$ 0.04& 0.14 $\pm$ 0.05& ...& ...& $162_{-50}^{+326}$ & $68_{-10}^{+12}$ & ARC \cr
HD 229059& 0.28 $\pm$ 0.09& 1.14 $\pm$ 0.17& 1.27 $\pm$ 0.20& 1.10 $\pm$ 0.03& 0.72 $\pm$ 0.06& 0.60 $\pm$ 0.05& 0.29 $\pm$ 0.03& 0.25 $\pm$ 0.03& 0.26 $\pm$ 0.04& ...& $58_{-4}^{+4}$ & $65_{-10}^{+12}$ & ARC \cr
CygOB2 5& 0.47 $\pm$ 0.06& 1.82 $\pm$ 0.09& 2.62 $\pm$ 0.15& 1.72 $\pm$ 0.10& 1.13 $\pm$ 0.11& 0.73 $\pm$ 0.10& 0.53 $\pm$ 0.10& ...& ...& ...& $84_{-16}^{+42}$ & $71_{-13}^{+15}$ & HIR \cr
CygOB2 12& 1.64 $\pm$ 0.03& 5.37 $\pm$ 0.04& 5.31 $\pm$ 0.13& 3.64 $\pm$ 0.05& 2.55 $\pm$ 0.05& 1.59 $\pm$ 0.04& 1.04 $\pm$ 0.04& 0.51 $\pm$ 0.04& ...& ...& $82_{-1}^{+2}$ & $40_{-3}^{+4}$ & ARC \cr
HD 203532& 0.22 $\pm$ 0.06& 0.98 $\pm$ 0.16& 0.89 $\pm$ 0.14& 0.60 $\pm$ 0.13& 0.39 $\pm$ 0.12& 0.45 $\pm$ 0.13& 0.30 $\pm$ 0.11& ...& ...& ...& $70_{-8}^{+8}$ & $37_{-8}^{+11}$ & HSF \cr
HD 203938& 0.21 $\pm$ 0.03& 0.50 $\pm$ 0.04& 0.70 $\pm$ 0.05& 0.63 $\pm$ 0.05& 0.55 $\pm$ 0.05& 0.24 $\pm$ 0.05& 0.47 $\pm$ 0.10& ...& ...& ...& $60_{-16}^{+412}$ & $88_{-34}^{+56}$ & ARC \cr
HD 204827& 1.76 $\pm$ 0.15& 5.63 $\pm$ 0.15& 5.06 $\pm$ 0.15& 3.38 $\pm$ 0.15& 1.80 $\pm$ 0.15& 0.78 $\pm$ 0.15& 0.64 $\pm$ 0.15& 0.51 $\pm$ 0.06& 0.24 $\pm$ 0.04& ...& $102_{-4}^{+6}$ & $38_{-4}^{+5}$ & ARC \cr
HD 206267& 0.41 $\pm$ 0.02& 1.23 $\pm$ 0.03& 1.07 $\pm$ 0.03& 0.58 $\pm$ 0.02& 0.31 $\pm$ 0.02& 0.19 $\pm$ 0.02& 0.11 $\pm$ 0.02& ...& ...& ...& $132_{-8}^{+8}$ & $35_{-3}^{+3}$ & STIS \cr
HD 207198& 0.21 $\pm$ 0.03& 0.73 $\pm$ 0.05& 0.88 $\pm$ 0.05& 0.69 $\pm$ 0.05& 0.52 $\pm$ 0.05& 0.38 $\pm$ 0.04& 0.21 $\pm$ 0.05& 0.17 $\pm$ 0.04& 0.13 $\pm$ 0.04& ...& $56_{-1}^{+2}$ & $59_{-5}^{+7}$ & S07 \cr
HD 207308& 0.35 $\pm$ 0.06& 1.15 $\pm$ 0.13& 1.48 $\pm$ 0.14& 0.81 $\pm$ 0.13& 0.50 $\pm$ 0.12& 0.47 $\pm$ 0.11& 0.41 $\pm$ 0.11& ...& ...& ...& $76_{-10}^{+14}$ & $45_{-12}^{+15}$ & HSF \cr
HD 207538& 0.37 $\pm$ 0.11& 1.58 $\pm$ 0.21& 1.54 $\pm$ 0.11& 1.32 $\pm$ 0.05& 0.80 $\pm$ 0.05& 0.72 $\pm$ 0.03& 0.83 $\pm$ 0.07& 0.45 $\pm$ 0.17& ...& ...& $52_{-8}^{+10}$ & $46_{-13}^{+14}$ & G06 \cr
HD 210121& 0.36 $\pm$ 0.05& 1.28 $\pm$ 0.08& 1.34 $\pm$ 0.08& 0.95 $\pm$ 0.08& 0.68 $\pm$ 0.08& 0.50 $\pm$ 0.07& 0.18 $\pm$ 0.06& ...& ...& ...& $82_{-6}^{+12}$ & $52_{-9}^{+11}$ & S07 \cr
HD 281159& 0.21 $\pm$ 0.02& 0.83 $\pm$ 0.03& 0.75 $\pm$ 0.03& 0.76 $\pm$ 0.03& 0.48 $\pm$ 0.03& 0.26 $\pm$ 0.03& 0.24 $\pm$ 0.03& ...& ...& ...& $72_{-10}^{+34}$ & $60_{-17}^{+23}$ & ARC \cr

\enddata
\tablenotetext{a}{References: ARC = ARC echelle data; G99
= Gredel 1999;  G06 = Galazutdinov et al. 2006; HIR = HIRES data; HSF = Hupe et al. 2012; 
IG11 = Iglesias-Groth 2011; K09 = Ka{\'z}mierczak et al. 2009; K10 = Ka{\'z}mierczak et al. 2010; 
S07 = Sonnentrucker et al. 2007; UV = UVES data; STIS = HST/STIS data;
vDB = van Dishoeck \& Black 1989; W20 = Welty et al. 2020}
\end{deluxetable}

In Figure 4, example results are shown for the sightline toward HD 24534.  Here, 
we present contours of $\chi^2$ in the
plane of gas density, $n_{\rm H}$, and kinetic temperature, $T$.  
Blue contours 
show the results obtained with the new NK20 collision rate coefficients, and red contours
are those obtained with the collisional cross-sections used by vDB82.  Blue and red dots 
show the best-fit values in each case, and the three contours of each color 
show the 1, 2, and 3 $\sigma$ error ellipses.  These were obtained by scaling the stated
observational errors to yield a minimum $\chi^2$ equal to the number of degrees of freedom, 
and then plotting contours of constant $\chi^2$ equal to that minimum value plus 
1, 4, and 9. The two fits are
equally good, with reduced $\chi^2$ values of 1.29 and 1.28 respectively, but the best-fit
densities differ by a factor of 7.8.

As expected given the collisional cross-sections shown in Figure 2, the derived gas
densities with the NK20 rate coefficients are dramatically lower than 
those obtained with the vDB82 estimates; the temperatures, by contrast, are almost
identical because they simply reflect the slope of the rotational diagram for small
$J$.  The blue points in Figure 5 show the
ratio of the densities obtained with the vDB82 estimates to those obtained with the 
NK20 rate coefficients.   Results are shown here as a function of the inferred
gas temperature (obtained with the vDB rates).

\includegraphics[angle=0,width=\textwidth]{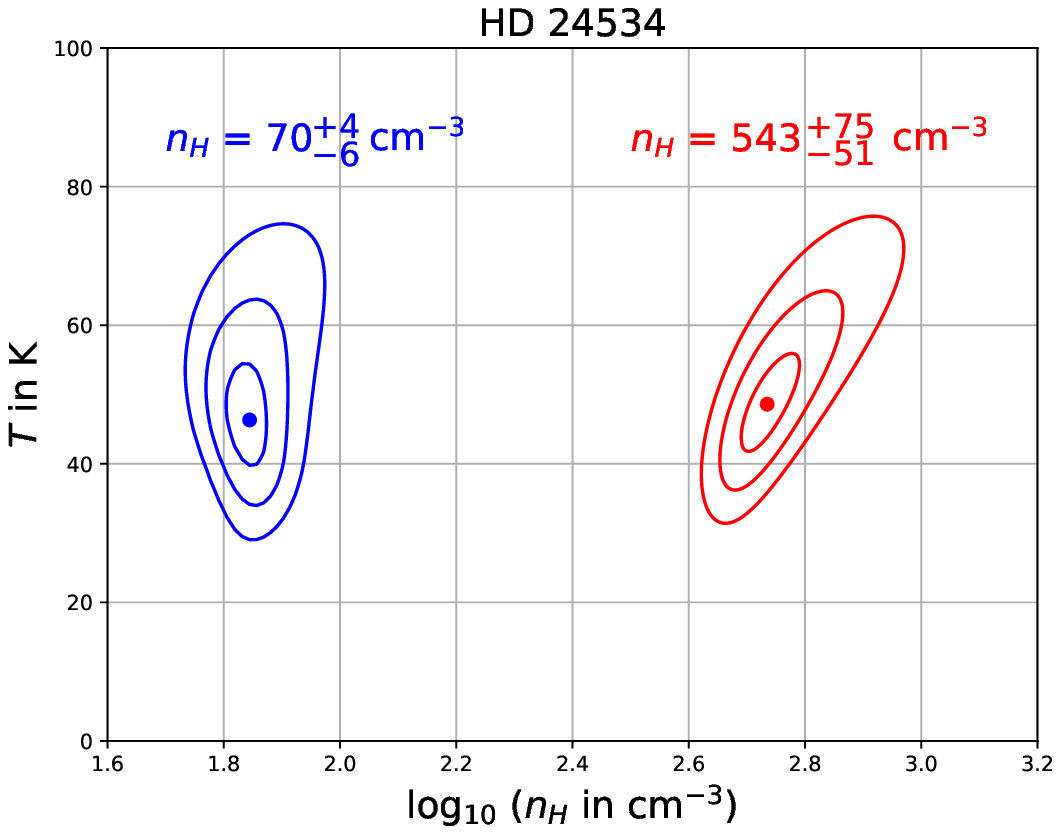}
\figcaption{Contours of constant $\chi^2$ in the
plane of gas density, $n_{\rm H}$, and kinetic temperature, $T$,
for the fit to the C$_2$ rotational diagram observed toward HD 24534.
Blue: 1$\sigma$, 2$\sigma$, and 3$\sigma$ error ellipses obtained with
the NK20 collisional rate coefficients.  Red: results obtained with the collisional 
cross-sections used by vDB82}

One caveat here is that the excitation is modeled for a medium of constant 
temperature and density.  There are indeed clear indications of multiple C$_2$ components, 
with somewhat different properties, for some sight lines (Sembach et al.\ 1996, CC12).
If an admixture of widely-varying temperatures were present 
along the sightline, then the rotational diagram could acquire some degree of curvature, 
even in the high density limit where collisional excitation dominates radiative pumping.
Widely-varying temperatures are not expected within diffuse molecular clouds that
are heated by interstellar UV radiation, but might exist if shocks are present.
\ree{More generally, the dissipation of turbulence can lead to additional heating terms
that increase the temperature beyond that expected from UV photoelectric heating alone, 
but the fraction of the gas so affected is likely too small (e.g. Godard et al.\ 2009)
to affect the C$_2$ rotational diagrams significantly.}

A second caveat concerns the radiation field near $1 \mu$m that is primarily responsible 
for radiative pumping of C$_2$.  The density estimates we obtain are proportional
to the radiative pumping rate that is assumed, $W_J$.  \re{The value adopted 
here is that expected given the mean ISRF assumed by
vDB82, scaled by a factor 1.39 to match the 1.25$\mu$m measurements performed by 
{\it COBE}/DIRBE.   Within diffuse molecular clouds, the radiation field might 
be reduced somewhat by attenuation by dust, although the effect is fairly minor 
at the relevant wavelengths, with the pumping rate decreasing with extinction along
each ray, $A_V$, as $\exp(-0.35\,A_V).$   Conversely, if the cloud is located close to a 
bright source of $1 \mu$m radiation, the radiation field could be enhanced.  A more
complete analysis, which we defer to a future paper, would involve computing the IR 
radiation at each point within an absorbing cloud, based upon the known three-dimensional
distribution of nearby stars and dust.  In the present study, we simply adopt the ISRF 
present at
the solar neighborhood, as calibrated using the all-sky 1.25$\mu$m map obtained by
DIRBE.}

\includegraphics[angle=0,width=5.5 true in]{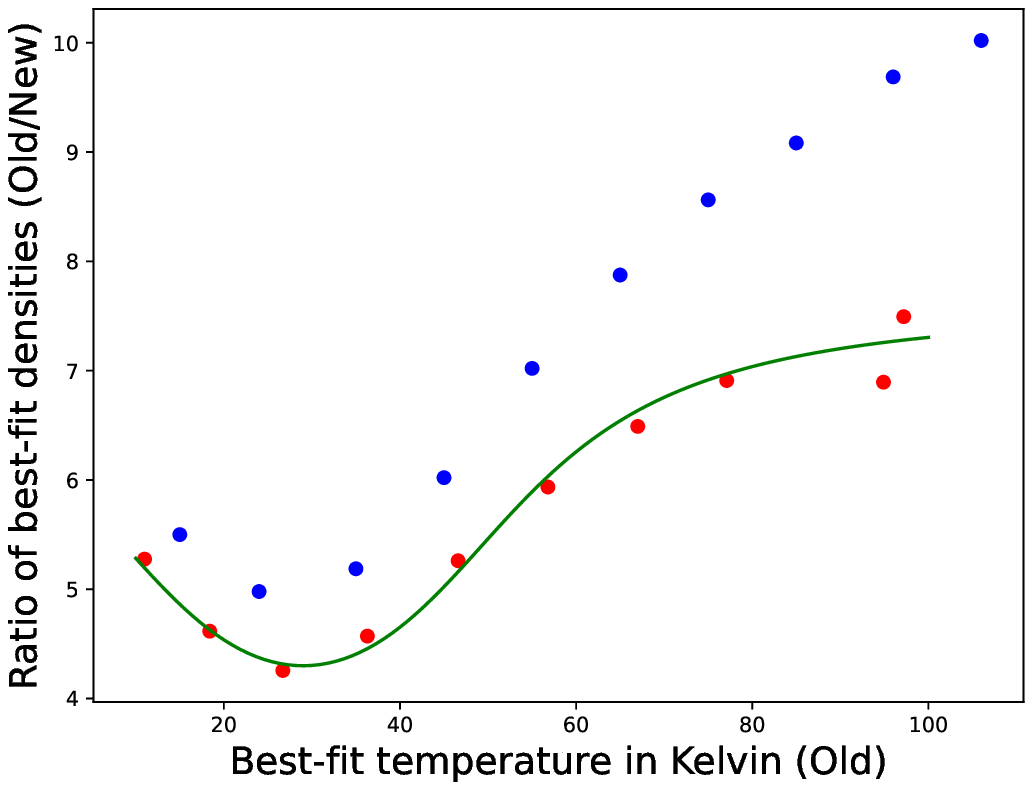}
\figcaption{Ratio of the gas densities obtained with the vDB82 
rate coefficients (blue points)
to those obtained with the NK20 rate coefficients; the same IR radiation field
and radiative excitation matrix are adopted in both cases. 
Red points show the ratio of the densities obtained with the 
OC to those obtained in the present work (i.e.\ with the
NK20 rate coefficients).
Results were obtained for rotational diagrams generated 
for $n_{\rm H}= 80\,\rm cm^{-3}$ with the collisional rates
calculated by NK20.  The green curve shows the fit to the
red points presented in eqn.\ (2), and may be used to correct
results obtained with the OC.}

\ree{A third caveat is that we assume hydrogen to be entirely molecular within
the region where C$_2$ is most abundant.  This approximation is supported by the 
observational finding that the C$_2$ 
column densities measured toward background stars, $N({\rm C_2})$, show a very strong 
dependence on the measured 
H$_2$ column densities, $N({\rm H_2})$ (e.g. S07).  This behavior is also 
consistent with chemical models (e.g.\ Federman \& Huntress 1989) that suggest 
the formation of C$_2$ is 
initiated by reaction of C$^+$ and CH, the latter molecule being known to be a 
surrogate tracer for H$_2$ (Sheffer et al.\ 2008).
In this context, it must be noted
that the local atomic fraction, $n({\rm H})/[2n({\rm H_2})+n({\rm H})]$, in the
densest regions where C$_2$ is most abundant, is expected to be 
considerably smaller than the ratio of column densities for the entire sightline, 
$N({\rm H})/[2N({\rm H_2})+N({\rm H})]$.  Among the 24 sources in our sample 
for which H and H$_2$ measurements are available, 
the latter quantity has a median value of 0.56; 
by contrast, 3D PDR models (Obolentseva et al.\ 2024) 
for selected sources predict a local atomic fraction $\sim 0.02 - 0.06$ for the hydrogen
in the densest material along the sightline (Obolentseva 2024, priv. comm.)}
\reee{
In other words, the gas composition along any given sightline may be very variable, 
with the local H abundance, $n({\rm H})/n_{\rm H}$, being reduced in the regions 
of highest density and anticorrelated with the C$_2$ abundance.  Most of the H$_2$ and 
C$_2$ column density along any diffuse cloud sightline is expected to come from 
the most highly shielded regions that surround the locations of maximum density.    
This behavior is invariably predicted by diffuse cloud models, dating back to the 
pioneering work of van Dishoeck \& Black (1986), and is a consequence of the non-linear 
behavior caused by H$_2$ self-shielding.  Because H$_2$ is photodissociated following line 
absorption in its ultraviolet Lyman and Werner bands, it is very effective in shielding 
itself once the line optical depths become significant (i.e once 
$N({\rm H}_2) \simgt 10^{13}\, \rm cm^{-2}$), 
leading to a very strong dependence of the molecular fraction on the shielding column, 
$N_{\rm H}$ (e.g. Dishoeck \& Black 1986; their Figure 3, which also shows a similarly strong 
dependence of the C$_2$ abundance).  Subsequent analytic treatments (Sternberg et al. 2024, 
and references therein) indicate that for the densities of present interest, the H$_2$/H 
transition occurs at a shielding column density that is only logarithmically dependent 
on gas density.   Thus, as shown by Obolentseva et al. (2024), H$_2$ self-shielding permits 
high molecular fractions to be achieved even at the lower gas densities inferred in our 
study.}

Many cloud density estimates in the literature (e.g. Fan et al. 2023) have been obtained 
using an online calculator\footnote{http://dib.uiuc.edu/c2/} (OC), developed by B.\ McCall, which fits any input set
of level populations, $f_J$, and uses the collisional cross-sections adopted
by vDB82.   The results generated by this online calculator are slightly different from
those that we obtain with the vDB82 collisional rates, because the OC uses an
earlier treatment of radiative pumping (due to vDB82 rather than CC12) and
a smaller assumed infrared radiation field.  We used the following method to estimate 
the correction factor needed to correct density estimates obtained from the OC 
or from our own treatment with the vDB82 collisional rates, with results that are
presented in Figure 5.
We first used the NK20 rates to compute the rotational diagrams 
expected for fully molecular gas with a density $n_{\rm H}= 80\,\rm cm^{-3}$ 
and a gas temperature ranging from 10 to 100~K in steps of 10~K; and we then fit those rotational diagrams using the 
vDB82 rates.  The best fit temperatures and densities are shown in Figure 5. 
Here, the red points show the values yielded by the OC, and the blue points
show those obtained with our own treatment using the vDB82 rates.  Both treatments
recover the temperatures to within 6 K at any temperature, but the densities are
overestimated by the factors indicated on the vertical axis.  

Almost identical results 
are obtained if gas densities, $n_{\rm H}$, of $40\,\rm cm^{-3}$ or 
$160\,\rm cm^{-3}$ are adopted in place of $80\,\rm cm^{-3}$.
For the OC, this factor is well fit by the expression
$$ {n({\rm OC}) \over n({\rm NK20})} = 
\re{4.3 + 3.3}\,
{{\rm tan}^{-1}(0.0014\,(T/{\rm K}-29)^2) \over \pi/2},\eqno(2)$$
(green curve) which may be used to obtain revised density estimates from 
literature estimates that
were obtained with the OC.  In cases where the fractional populations have also been
published -- which was not the case for Fan et al.\ (2023), however -- a revised 
calculator is available as a python script upon request to the corresponding author.

While the new rate coefficients reduce the best-fit densities by a 
factor 4 -- 7 below the values obtained with the OC, the 
quality of the fits is indistinguishable.  In other words, the
fractional C$_2$ rotational populations, $f_J$, can be fit 
equally well for either
set of adopted rate cofficients. 
Figure 6 shows the 
cumulative distribution functions for the minimum reduced $\chi^2$,
with red and blue histograms applying to fits obtained with
the vDB82 and NK20 rates, respectively.  These distribution
functions apply to the set of 46 rotational diagrams that we fit. 
In roughly 30$\%$ of cases, the minimum reduced-$\chi^2$ was 
less than unity, implying that the stated error bars on the C$_2$ column 
densities were overestimates.  For the remaining 70$\%$ of cases that
yield a minimum reduced $\chi^2$ greater than unity, either 
(1) the errors bars were underestimated; or (2) our simple 
constant-density constant-temperature excitation models fail to account 
fully for the observed rotational diagrams.  In more than 
three-quarters of the 46 cases, the minimum reduced-$\chi^2$ lies
within the range 0.25 -- 4.  If the deviations from unity are
the result of inaccuracies in the error bars (within an 
inhomogeneous set of reported observations), then the subset with 
minimum reduced $\chi^2$ in the range 0.25 -- 4 have error bars that 
are correct to within a factor 2.

\includegraphics[angle=0,width=5.5 true in]{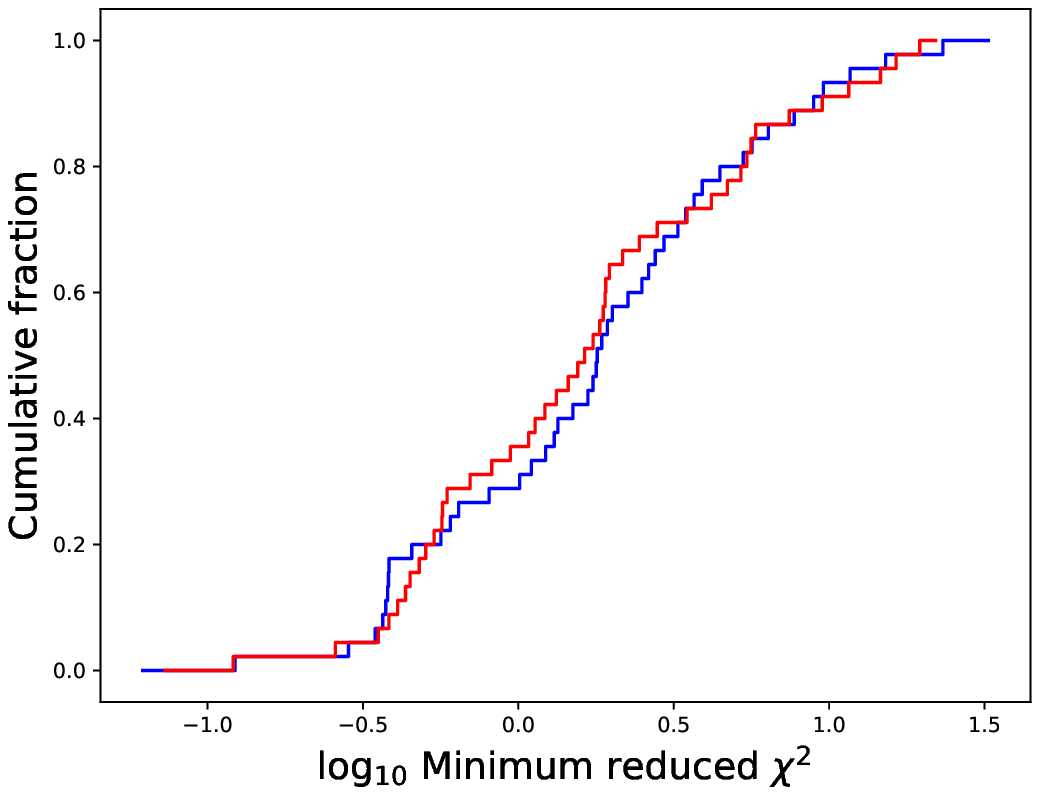}[b]
\figcaption{Cumulative distribution functions for the minimum reduced $\chi^2$,
with red and blue histograms applying to fits obtained with
the vDB82 and NK20 rates, respectively}

The gas temperature and density estimates obtained with the CC12 treatment of radiative 
excitation and the NK20 collisional rates are presented in Table 1, along with
the observed column densities, $N(J)$, in state $J$.  In Figure 7, we
present a compilation of the best-fit results and 1-$\sigma$ error bars in the
$n({\rm H_2}) - T$ plane.   The black, magenta, blue, cyan, green and red curves show contours of
constant gas pressure with $p/k =$ 1000, 1500, 2000, 2500, 3000, and 3500~$\rm K\,cm^{-3}$, respectively;
here, we assume that the gas probed by C$_2$ is almost fully molecular and we include
the contribution of He to the pressure.  The mean and standard deviation of the best-fit ${\rm 
log}_{10}(p/k)$ estimates are 3.35 and 0.19, respectively -- values that are in reasonable agreement
with those  obtained for the diffuse ISM by Jenkins \& Tripp (2011) 
from observations of [CI] absorption (3.58 and 0.175) -- although for 
individual sightlines where both [CI] and C$_2$ are observed, the pressure estimates may
differ by up to a factor 20 in rare cases.

\includegraphics[angle=0,width=5 true in]{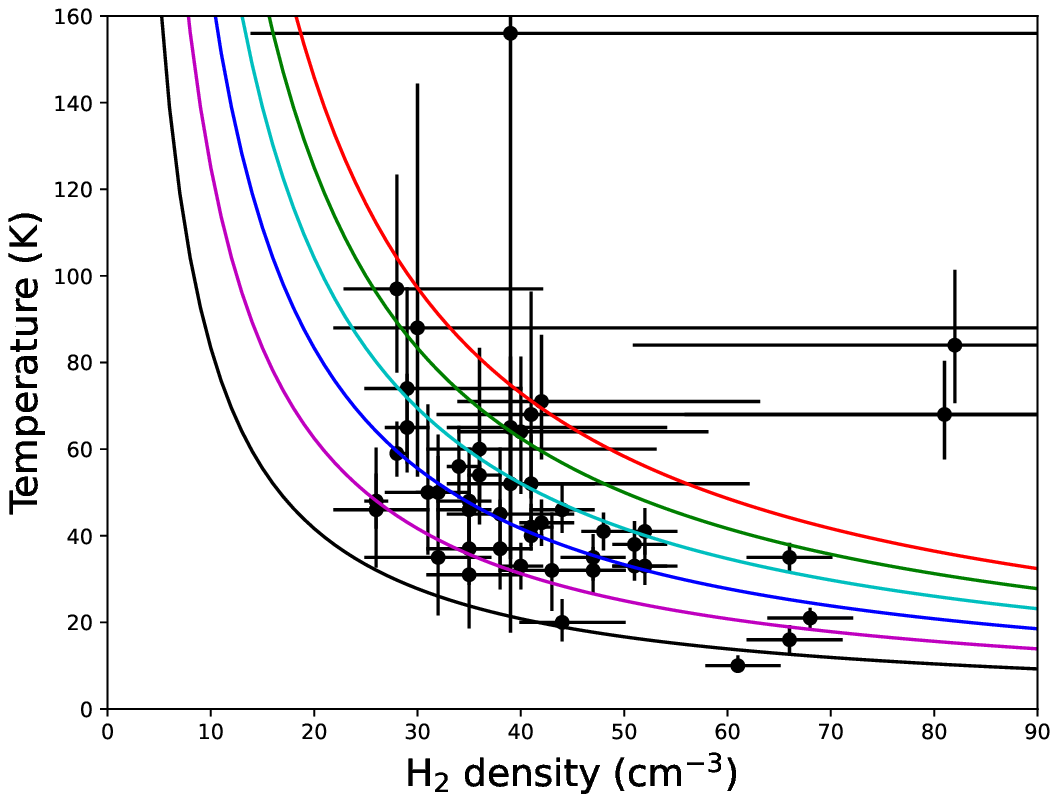}
\figcaption{Best-fit results and 1-$\sigma$ error bars in the
$n({\rm H_2}) - T$ plane, for our sample of 46 background sources (Table 1).
The black, magenta, blue, cyan, green and red curves show contours of
constant gas pressure with 
$p/k =$ 1000, 1500, 2000, 2500, 3000, and 3500~$\rm K\,cm^{-3}$, respectively.}

\section{Comparison with densities determined from 3-dimensional extinction maps}

To perform a comparison with the C$_2$-derived density estimates, we considered those
23 sources listed in Table 1 that have a {\it Gaia}-determined
distance, $d_{\rm star}$, within 1.25~kpc of the Sun (i.e. within the volume covered by the E23 extinction map).
These are listed in Table 2.
For each star, we computed the peak foreground density along the sightline implied by both the 
E23 and L20 extinction maps, as well as the heliocentric distances at which the 
density is maximal, $d_{\rm cloud}({\rm E23})$ and $d_{\rm cloud}({\rm L20})$.  
\re{In cases where it is not entirely clear whether $d_{\rm cloud}$ is smaller or larger
than $d_{\rm star}$, we adopt a peak foreground density under the assumption that the
star lies {\it behind} the cloud; in these cases, the adopted values may be overestimates.}
  
In Figure 8, we show the density 
implied by the E23 extinction map within a plane of constant
Galactic longitude, $l$, containing the line-of-sight to HD 24534.  
The vertical axis shows the distance in the $Z$-direction 
(perpendicular to the Galactic disk), and the horizontal axis 
shows the distance in the orthogonal direction within the slice of
constant $l$.  The Sun is located at the origin within each panel, 
and the dashed curve is the line-of-sight.  The top left panel shows
the entire region covered by E23, with the location of the
background source marked with a cross.  The other
panels show successively zoomed regions centered on the 
Sun (top right) and the location along the sightline where the
density is largest (bottom left).  The E23 extinction maps cover
heliocentric distances in the 69 -- 1250 pc range, and the 
unmapped regions appear in white.  As discussed in detail in E23, they are obtained from 
a Bayesian analysis of the available data. 
Twelve possible realizations of the extinction
density -- known as posterior estimates -- are provided by E23, each 
of which may be regarded as equally probable.
The densities plotted in Figure 8
are averages over these posterior estimates.  Here, we adopted a column density
conversion of \re{\bf $N_{\rm H} = 5.3 \times 10^{21}\, {\rm cm}^{-2} E_{\rm ZGR}$}, where $E_{\rm ZGR}$ is
the extinction defined by Zhang et al.\ (2023; hereafter ZGR), or equivalently
\re{$n_{\rm H} = 1.7 \times 10^3\, {\rm cm}^{-3} \times dE_{\rm ZGR}/ds$} where $s$ is the distance along 
the sightline in pc.  

\ree{This conversion is based upon the universal 
extinction curve given by ZGR, which implies a V-band extinction, $A_V=2.8\,E_{ZGR}$;
a ratio of $A_V$, to color excess, $E(B-V)$, of 3.1; and a ratio  
$N_{\rm H}$ to $E(B-V)$ of $5.8 \times 10^{21} \rm cm^{-2}$ (Bohlin et al.\ 1978).  
The latter two ratios imply that $N_{\rm H}/A_V = 1.9 \times 10^{21} \rm cm^{-2}.$
This value for $N_{\rm H}/A_V$ is in acceptable agreement with the mean value obtained 
subsequently by
Rachford et al.\ (2009) for a sample of 17 sightlines: 
$(2.15 \pm 0.14) \times 10^{21} \rm cm^{-2}$.  That
sample (which is smaller than that presented by Bohlin et al.\ 1978) 
comprises the set of available non-Be stars with reliable 
$N({\rm H}_2)$ and $N({\rm H})$ obtained through ultraviolet observations.
Variations about the best-fit $N_{\rm H}/A_V$ value for this sample (Rachford 
et al. 2009; see lower panel of their Figure 3) are barely larger than those 
expected from quoted measurement uncertainties.  Assuming that the latter are correctly
estimated, we find that the real rms sightline-to-sightline 
variation of $N_{\rm H}/A_V$ is only 12$\%$.  Naturally, any inaccuracies in the 
approximations discussed above would lead to inaccuracies in the inferred gas density, 
with the derived values of $n_{\rm H}$ being inversely proportional to the 
(G-band) extinction-to-gas ratio.}

\includegraphics[angle=0,width=\textwidth]{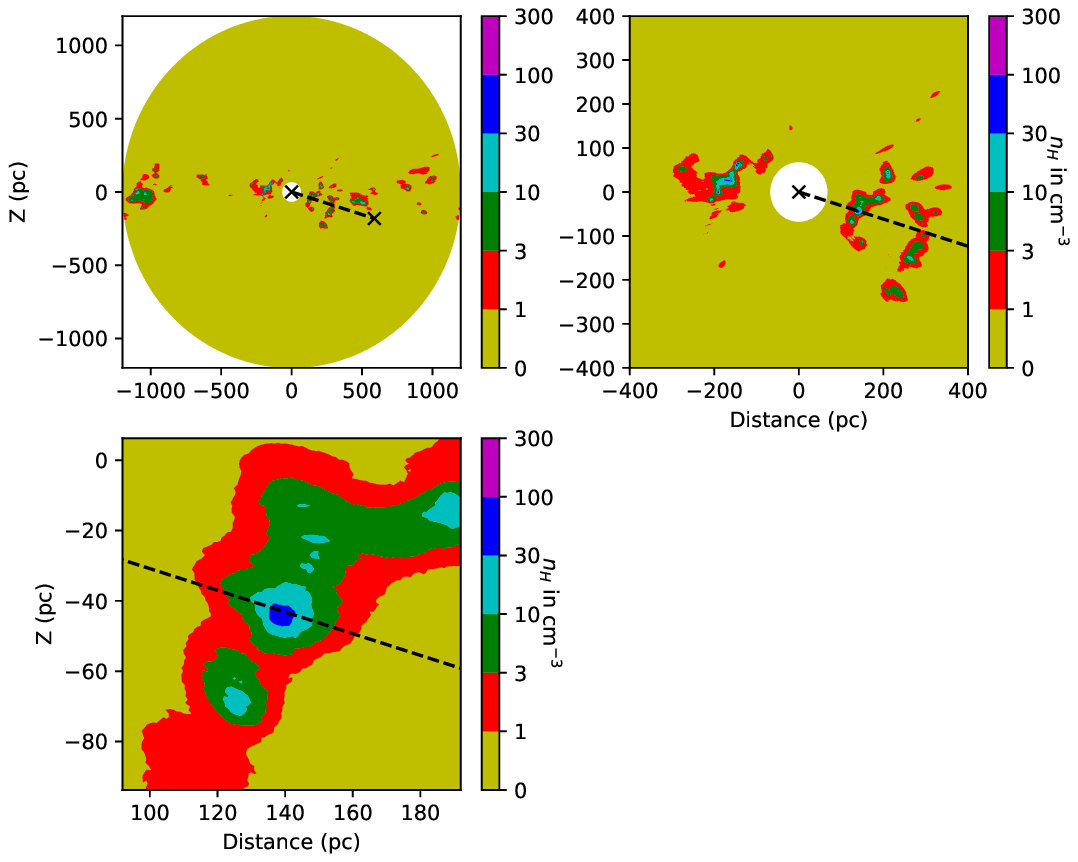}
\figcaption{Gas density, $n_{\rm H}$, 
implied by the E23 extinction map within a plane of constant
Galactic longitude, $l$, containing the line-of-sight to HD 24534.  }

Figure 9 shows the density, $n_{\rm H}$, averaged over the 12 posterior samples, 
as a function of distance along the sightline to HD 24534.  The upper panel 
shows the \re{density for the full sightline, using logarithmic spacing on vertical axis and} with the 
location of the star indicated by the red cross.  The lower panel zooms in on the
region of maximum density, showing the mean density estimate in black and those from two 
individual samples in red and blue.  These samples have the 2nd lowest and 2nd highest
peak densities among the 12 posterior samples.  We adopt the mean peak density within
the 12 samples as our best estimate of the peak density along the sightline.
As our lower and upper bounds, we adopt the peak values in the samples shown in red and
blue.  Two-thirds of the samples have peak density estimates lying between these two values
and one-sixth of the samples have peak density estimates lying outside these two values 
(with the remaining one-sixth having peak density estimates equal to these two values).  
We may therefore regard these values as corresponding to 75$\%$ confidence limits, 
75$\%$ being the mean of the fraction of peak density estimates lying inside the 
confidence limits (67$\%$) and the fraction that are not outside the limits (83$\%$).
These peak density estimates are tabulated in Table 2.

\includegraphics[angle=0,width=4.5 true in]{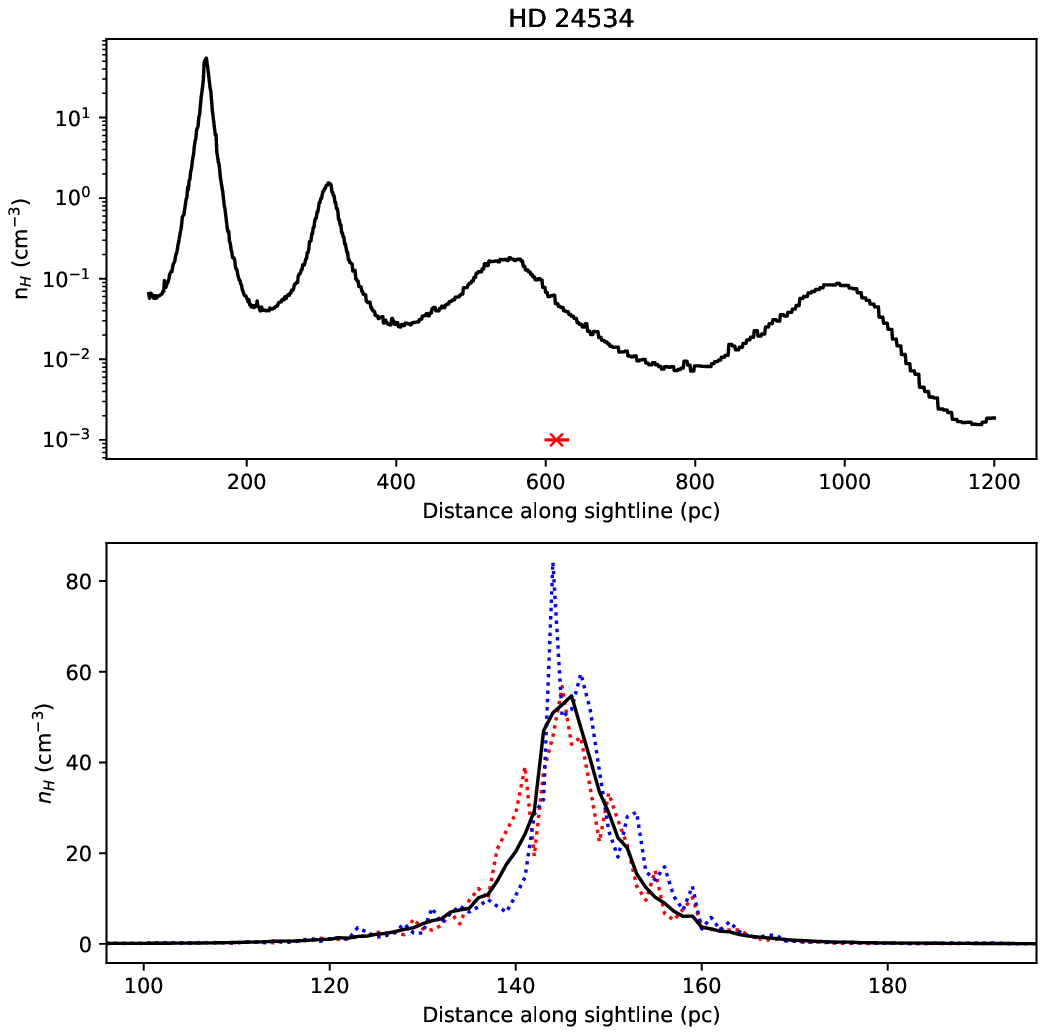}
\figcaption{Gas density, $n_{\rm H}$, 
averaged over the 12 posterior samples, 
as a function of distance along the sightline to HD 24534.
Upper panel: ${\rm log}_{10}(n_{\rm H}/{\rm cm}^{-3})$ for the full sightline.
Lower panel: $n_{\rm H}/{\rm cm}^{-3}$ near the 
region of maximum density, showing the mean density estimate in black and those from two 
individual samples in red and blue (see text).}

\begin{deluxetable}{lrrllllll}
\tablenum{2}
\tabletypesize{\scriptsize}
\tablecaption{Cloud density and distance estimates}
\tablehead{\cr
Source & $l\phantom{000}$ & $b\phantom{000}$ & $n_{\rm H}$ & $n_{\rm H}$ & $n_{\rm H}\,^a$
 & $d_{\rm cloud}$ &  $d_{\rm cloud}\,^a$ & $d_{\rm star}$ \cr
 &  &       & ${\rm C}_2$ & {\rm E23} & {\rm L20} & {\rm E23} & {\rm L20} & {\rm {\it Gaia}$^b$} \cr
 &   (deg)    &    (deg)        &    (cm$^{-3}$)         & (cm$^{-3}$)& (cm$^{-3}$)& (pc) & (pc) & (pc)}
\startdata
HD 24534& 163.0814& $-17.1362$& $70_{-6}^{+4}$ & $71_{-14}^{+13}$ & $101_{-34}^{+43}$ & 146& 148& $614_{-14}^{+14}$ \cr
HD 27778& 172.7629& $-17.3928$& $64_{-4}^{+2}$ & $34_{-6}^{+5}$ & $45_{-16}^{+17}$ & 153& 138& $212_{-1}^{+1}$ \cr
HD 29647& 174.0529& $-13.3487$& $122_{-6}^{+8}$ & $103_{-31}^{+34}$ & $263_{-62}^{+97}$ & 156& 140& $156_{-1}^{+1}$ \cr
HD 34078& 172.0813& $-2.2592$& $56_{-10}^{+28}$ & $9_{-1}^{+1}$ & .....& 374&.....& $389_{-5}^{+5}$ \cr
HD 62542& 255.9153& $-9.2371$& $104_{-4}^{+6}$ & $13_{-3}^{+3}$ & $313_{-47}^{+73}$ & 294& 295& $367_{-2}^{+2}$ \cr
HD 73882& 260.1816& $0.6431$& $132_{-8}^{+10}$ & $17_{-4}^{+6}$ & .....& 899&.....& $461_{-136}^{+333}$ \cr
HD 147084& 352.3279& $18.0503$& $62_{-8}^{+8}$ & $244_{-36}^{+34}$ & $172_{-58}^{+74}$ & 150& 132& $270_{-34}^{+46}$ \cr
HD 147888& 353.6470& $17.7092$& $68_{-2}^{+4}$ & $230_{-41}^{+39}$ & $189_{-69}^{+74}$ & 150& 133& $124_{-6}^{+6}$ \cr
HD 147889& 352.8573& $17.0436$& $72_{-2}^{+4}$ & $189_{-19}^{+25}$ & $149_{-54}^{+124}$ & 150& 132& $136_{-0}^{+0}$ \cr
HD 147933& 353.6860& $17.6867$& $84_{-4}^{+6}$ & $225_{-39}^{+40}$ & $184_{-68}^{+72}$ & 150& 136& $138_{-2}^{+3}$ \cr
HD 148184& 357.9328& $20.6766$& $78_{-12}^{+30}$ & $73_{-11}^{+11}$ & $53_{-23}^{+22}$ & 150& 117& $153_{-4}^{+5}$ \cr
HD 149757& 6.2812& $23.5877$& $52_{-2}^{+2}$ & $45_{-9}^{+12}$ & $42_{-14}^{+20}$ & 107& 105& $112_{-2}^{+3}$ \cr
HD 154368& 349.9702& $3.2151$& $80_{-4}^{+4}$ & $93_{-19}^{+16}$ & $75_{-23}^{+28}$ & 201& 200& $1064_{-37}^{+40}$ \cr
HD 170740& 21.0574& $-0.5259$& $70_{-8}^{+10}$ & $50_{-12}^{+16}$ & $35_{-17}^{+15}$ & 228& 226& $226_{-4}^{+5}$ \cr
HD 179406& 28.2285& $-8.3118$& $162_{-50}^{+326}$ & $19_{-4}^{+4}$ & $43_{-24}^{+46}$ & 227& 225& $289_{-9}^{+10}$ \cr
HD 203532& 309.4590& $-31.7397$& $70_{-8}^{+8}$ & $43_{-11}^{+12}$ & $92_{-42}^{+61}$ & 205& 206& $291_{-2}^{+2}$ \cr
HD 204827& 99.1667& $5.5525$& $102_{-4}^{+6}$ & $14_{-3}^{+3}$ & .....& 461&.....& $929_{-78}^{+94}$ \cr
HD 206267& 99.2904& $3.7383$& $132_{-8}^{+8}$ & $7_{-1}^{+1}$ & .....& 461&.....& $735_{-102}^{+141}$ \cr
HD 207198& 103.1363& $6.9949$& $56_{-1}^{+2}$ & $11_{-2}^{+2}$ & .....& 426&.....& $1002_{-33}^{+35}$ \cr
HD 207308& 103.1090& $6.8176$& $76_{-10}^{+14}$ & $8_{-2}^{+1}$ & .....& 426&.....& $924_{-14}^{+14}$ \cr
HD 207538& 101.5990& $4.6727$& $52_{-8}^{+10}$ & $17_{-5}^{+4}$ & .....& 458&.....& $849_{-12}^{+12}$ \cr
HD 210121& 56.8751& $-44.4610$& $82_{-6}^{+12}$ & $37_{-13}^{+12}$ & $29_{-9}^{+8}$ & 239& 240& $334_{-4}^{+4}$ \cr
HD 281159& 160.4908& $-17.8022$& $72_{-10}^{+34}$ & $95_{-20}^{+28}$ & $115_{-30}^{+50}$ & 307& 302& $152_{-58}^{+250}$ \cr

\enddata
\tablenotetext{a}{Where $d_{\rm cloud}({\rm E23})$ lies outside the volume covered
by the L20 maps, no entry is given.}
\tablenotetext{b}{Gaia Collaboration (2020)}
\end{deluxetable}

In a similar manner, we obtained peak density estimates for the other 22 background sources lying
within 1.25 kpc of the Sun.  In Figure Set 1 (Appendix A), plots similar to Figure 1, Figure 4, and the 
bottom panels of Figures 8 and 9 are presented in a single figure for each source.
An example is shown in Figure 10, where we present the results for a more distant cloud, 
located along the sightline to HD 204827.

\includegraphics[angle=0,width=6 true in]{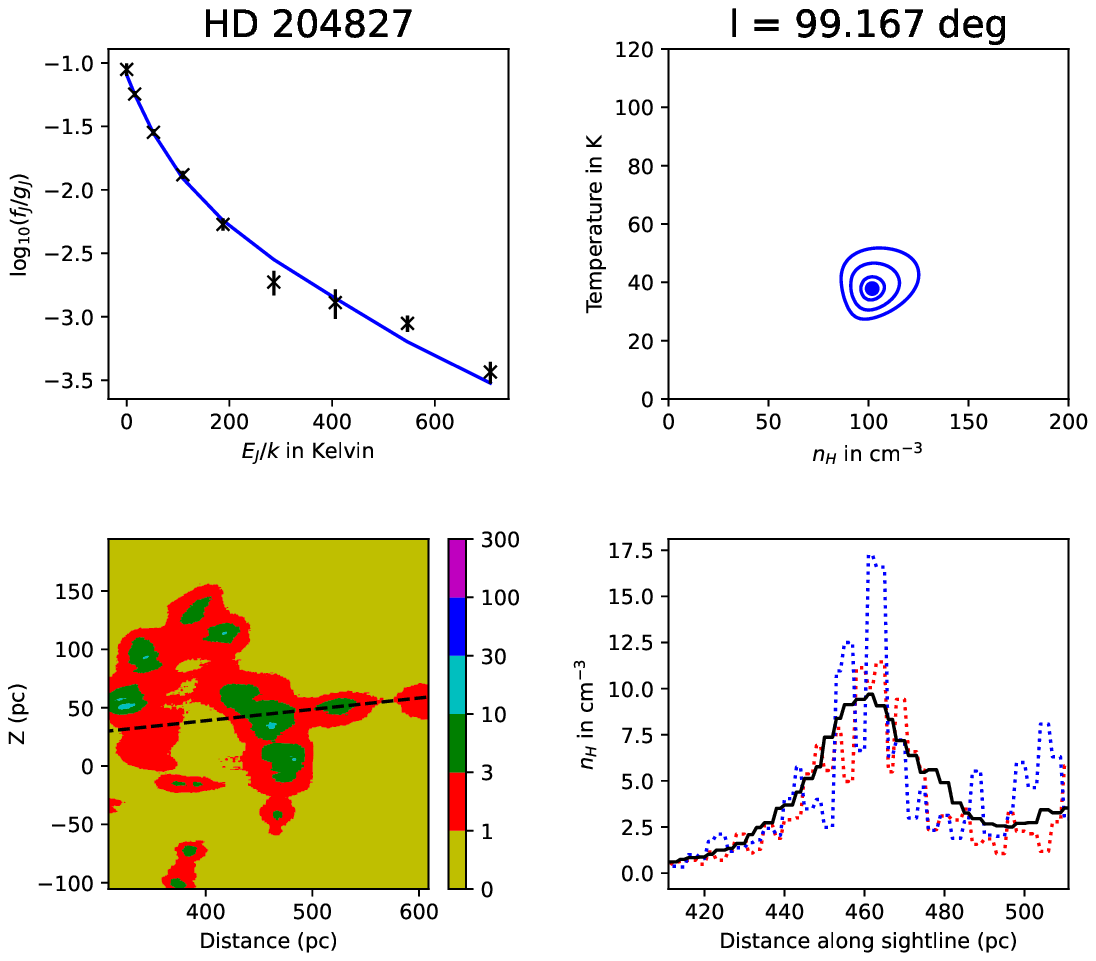}
\figcaption{Summary of results for the sightline to HD 204827.  The complete figure set 
(23 images, one for each sightline) is available in the online journal.
Top left: Rotational diagram from 
C$_2$ absorption line observations.  
Top right: Contours of constant $\chi^2$ in the
plane of gas density, $n_{\rm H}$, and kinetic temperature, $T$,
for the fit to the C$_2$ rotational diagram; here we show
1$\sigma$, 2$\sigma$, and 3$\sigma$ error ellipses obtained with
the NK20 collisional rate coefficients.  Bottom left: Gas density, $n_{\rm H}$, 
implied by the E23 extinction map within a plane of constant
Galactic longitude, $l$, containing the line-of-sight to the source. 
Bottom right: gas density, $n_{\rm H}$, 
averaged over the 12 posterior samples from E23, 
as a function of distance along the sightline (black).
Results from two individual samples are shown in red and blue (see text).} 

For the 17 sources where the position of peak density lies
within the (smaller) volume covered in the L20 map, we used entirely analogous methods to compute
the peak density given by the L20 map.  The comparison between the peak line-of-sight
density estimates in the two maps is shown in \ree{the top left panel of} Figure 11.   Most of the estimates agree 
within the \re{75$\%$ confidence limits obtained using the method described above.  These
errors, however, are statistical in nature and do not capture systematic uncertainties
in the model (see E23's Section 6).}

\includegraphics[angle=0,width=3.5 true in]{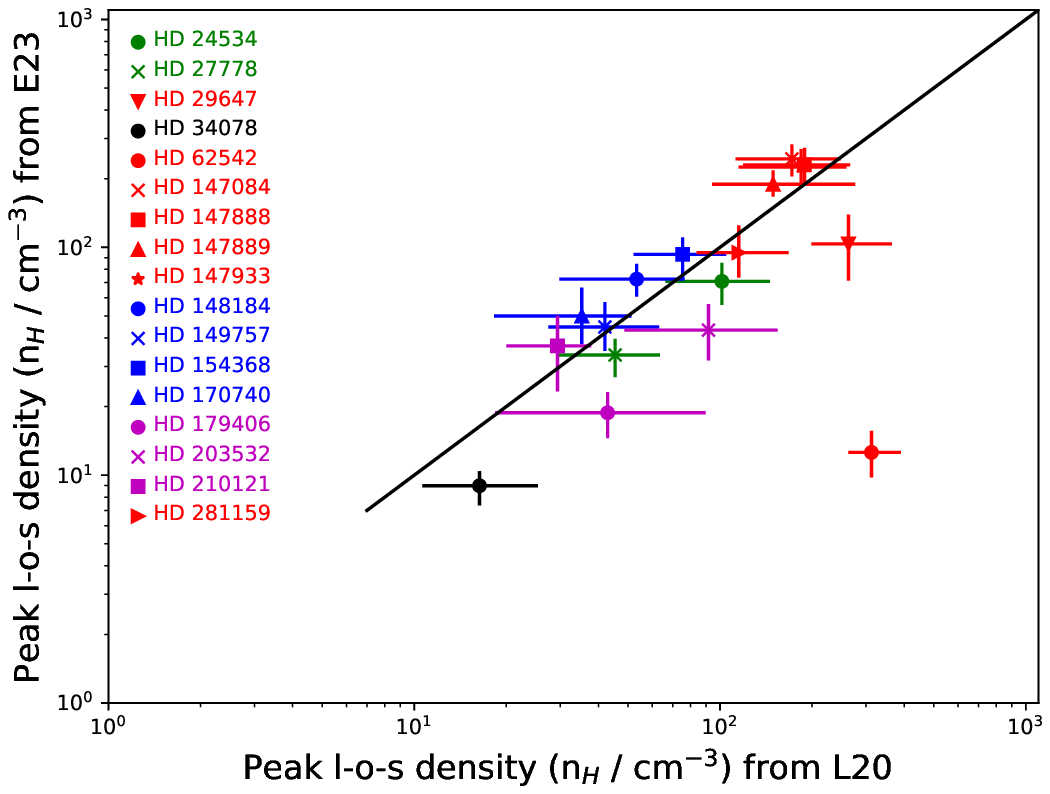} 
\includegraphics[angle=0,width=3.5 true in]{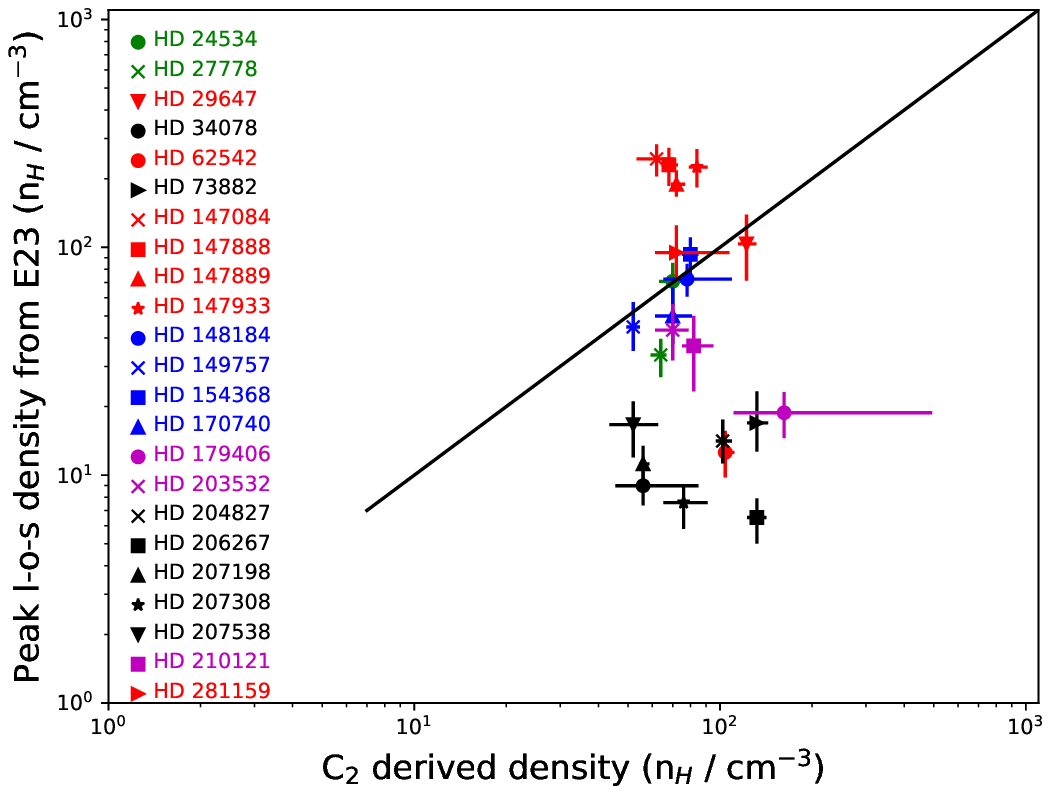} \\
\includegraphics[angle=0,width=5 true in]{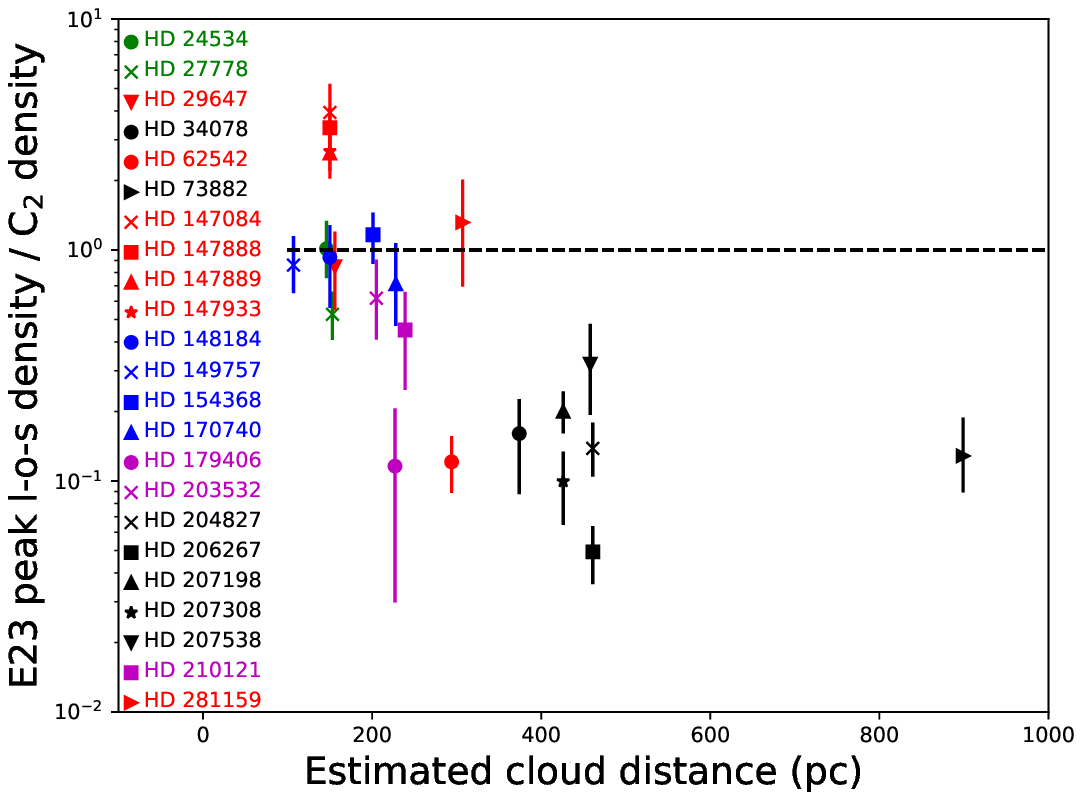} 
\figcaption{\ree{Top left: comparison between the peak line-of-sight
density estimates in the L20 and E23 extinction maps, for 17 sources.
Top right: comparison between
the peak densities obtained from the E23 maps with the densities derived
from C$_2$. Bottom: ratio of E23 extinction map density estimate to C$_2$ density estimate as
a function of the estimated cloud distance.}}

The red circle, applying to the sightline 
toward HD 62542, is an outlier;
this is a rare region where the resolution in the plane-of-the sky 
is insufficient to yield reliable 
density estimates for the small dust cloud that HD 62542 is likely associated with.
A dearth of background
stars in the vicinity of $l \sim 352^o$, $b \sim 18^o$ makes the extinction maps unreliable
for four additional sightlines (to HD~147084, HD~147888, HD~147889, and HD~147993). 
Similarly, HD 29647 and HD 281159 lie behind regions where the density of stars is insufficient
to yield reliable results.
These six cases are also plotted with red symbols in Figure 11. 

In the \ree{top right} of Figure 11, we compare the peak densities 
obtained from the E23 maps with the densities derived
from C$_2$.  \ree{C$_2$ is expected to probe the densest material 
along the sightline because
it is formed via a reaction sequence initiated by the reaction of C$^+$ with CH (Federman and
Huntress 1989) and CH is most abundant where the gas is denser and more molecular}. 
The black line at 45 degrees indicates where the two density estimates are equal.  At least
in an average sense, the new NK20 collisional rates bring the C$_2$ density estimates into reasonable
agreement with the peak densities inferred from the E23 extinction maps: the median best-fit values 
for the two estimates are $n_{\rm H} = 72\,\rm cm^{-3}$ and $44\,\rm cm^{-3}$, 
respectively, and mean best-fit values are $n_{\rm H} = 81\,\rm cm^{-3}$ and $74\,\rm cm^{-3}$.

\re{The four points (red symbols) lying significantly above the black line are precisely those in
the vicinity of $l \sim 352^o$, $b \sim 18^o$, where a dearth of background stars
makes the extinction maps unreliable.}

{Of the eight points lying significantly below
the line, one applies to the sightline HD 62542 (red circle) 
where the extinction map estimate is also known
to be unreliable.  Six of the remaining seven points represent the most distant density peaks
in the sample (all lying more than 350 pc from the Sun -- see Table 2).  These cases
are represented by black symbols.}  Since the spatial resolution
of the extinction maps worsens with increasing heliocentric distance, we speculate that these may be 
cases where the density peak is smeared out by inadequate resolution along the sightline.
\re{This speculation is supported by the map shown in Figure 10 (bottom left panel),
in which the gas exhibits artificial extensions 
(``fingers of God'') pointing towards the Sun.}
\ree{The bottom panel of Figure 11} shows the ratio of extinction map density estimate to C$_2$ density estimate as
a function of the estimated cloud distance.

\re{For the 11 sightlines with (1) sufficient background stars to yield reliable extinction map 
density estimates and (2) where the density peak lies within 350 pc of the Sun, the best-fit
C$_2$ density estimates agree within a factor 2 in all but one case.  For this subset of
the sightlines, we computed the mean density with a weighting inversely proportional to
the squares of the uncertainties.  The values obtained from the extinction maps and from the
C$_2$ observations are $n_{\rm H} = 56\,\rm cm^{-3}$ and $65\,\rm cm^{-3}$ respectively, 
values that agree to within a factor better than 1.2.}

\section{Discussion}

Our significant downwards revision in the cloud density estimates derived from
observations of C$_2$ -- which is corroborated by the extinction maps as discussed
in Section 3 -- has implications for our understanding of the diffuse ISM.  In particular,
C$_2$ density estimates have been widely used in deriving the cosmic-ray ionization
rate (CRIR) from absorption line observations of the H$_3^+$ molecular ion
(McCall et al.\ 2003, Indriolo et al.\ 2007, Indriolo \& McCall 2012).  
Density estimates
are essential for this purpose because the H$_3^+$ molecular ion is formed at a rate per
H$_2$ molecule that is dependent on the CRIR and destroyed at a rate that is proportional
to the gas density.  The abundance of H$_3^+$ relative to H$_2$ is therefore dependent
on the ratio of the CRIR to the gas density.  The derived CRIR, in turn, depends on
what is assumed for the gas density and will accordingly be reduced significantly.  In our
earlier study of the CRIR (Neufeld and Wolfire 2017; hereafter NW17), 
we obtained an average estimate of the primary ionization 
rate per H atom, $\zeta_p({\rm H})$, of $(2.3 \pm 0.6) \times 10^{-16}\,\rm s^{-1}$.
The error given here is simply the standard error on the mean, without inclusion
of possible systematic effects, and the value is the mean obtained for 7 ``gold-plated"
sightlines for which direct absorption line observations were available for H$_2$, C$_2$
and H$_3^+$ (at ultraviolet, near-infrared and mid-infrared wavelengths respectively).
The C$_2$ density estimates adopted were literature values that had been obtained with 
the online calculator.

Of the 7 sightlines with absorption line observations for H$_2$, C$_2$, and H$_3^+$,
four have C$_2$ observations covering $J \ge 12$ and are therefore in the sample adopted
in the present work (HD 24534, HD 27778, HD 73882, and HD 154368).   Using the revised 
NK20 rate coefficients to obtain
density estimates for these sightlines, we used the diffuse cloud models of NW17 to
obtain estimates of the CRIR from the observed H$_3^+$ abundances: the mean value 
was \re{\bf $\zeta_p({\rm H})= 8 \times 10^{-17}\,\rm s^{-1},$} corresponding to a total 
ionization rate
per H$_2$ molecule of $\zeta_t({\rm H_2})= 1.8 \times 10^{-16}\,\rm s^{-1}.$  
The actual CRIR may be even lower, because 
C$_2$ is likely weighted toward denser, more fully-molecular gas than H$_3^+$.

With the availability of the 3D density maps derived from E23 and L20 --
and validated
to some extent in Section 3 above -- it is now possible
to construct more sophisticated models than the simple constant density slab models adopted 
by NW17.  In a companion paper by Obolentseva et al.\ (2024), such models are used to
obtain predictions of the H$_3^+$ density in 3 dimensions, as a function of 
the CRIR.  These can then be compared with the H$_3^+$ column densities observed 
along each sightline, thereby providing estimates of the CRIR and how it varies. 

We are very grateful to F.\ Najar for providing full results from NK20 in
machine-readable tables. 

\newpage
\appendix
\centerline{\bf Figure Set 1}
\centerline{(see Figure 10 for full caption)}
\begin{figure}[b]
\includegraphics[angle=0,width=6 true in]{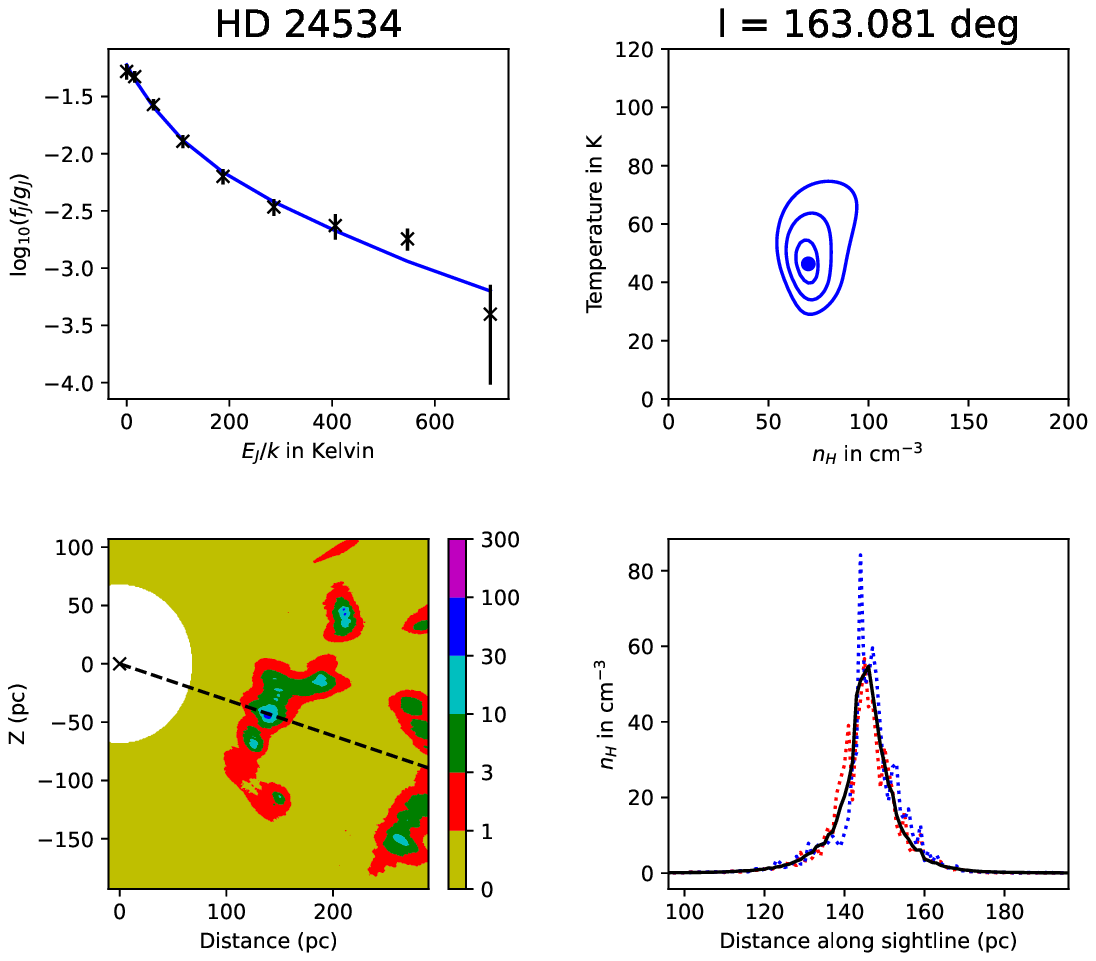}
\figurenum{10.1}
\figcaption{Results for the line-of-sight to HD 24534}
\vskip 2 true in
\end{figure}

\begin{figure}[!]
\includegraphics[angle=0,width=6 true in]{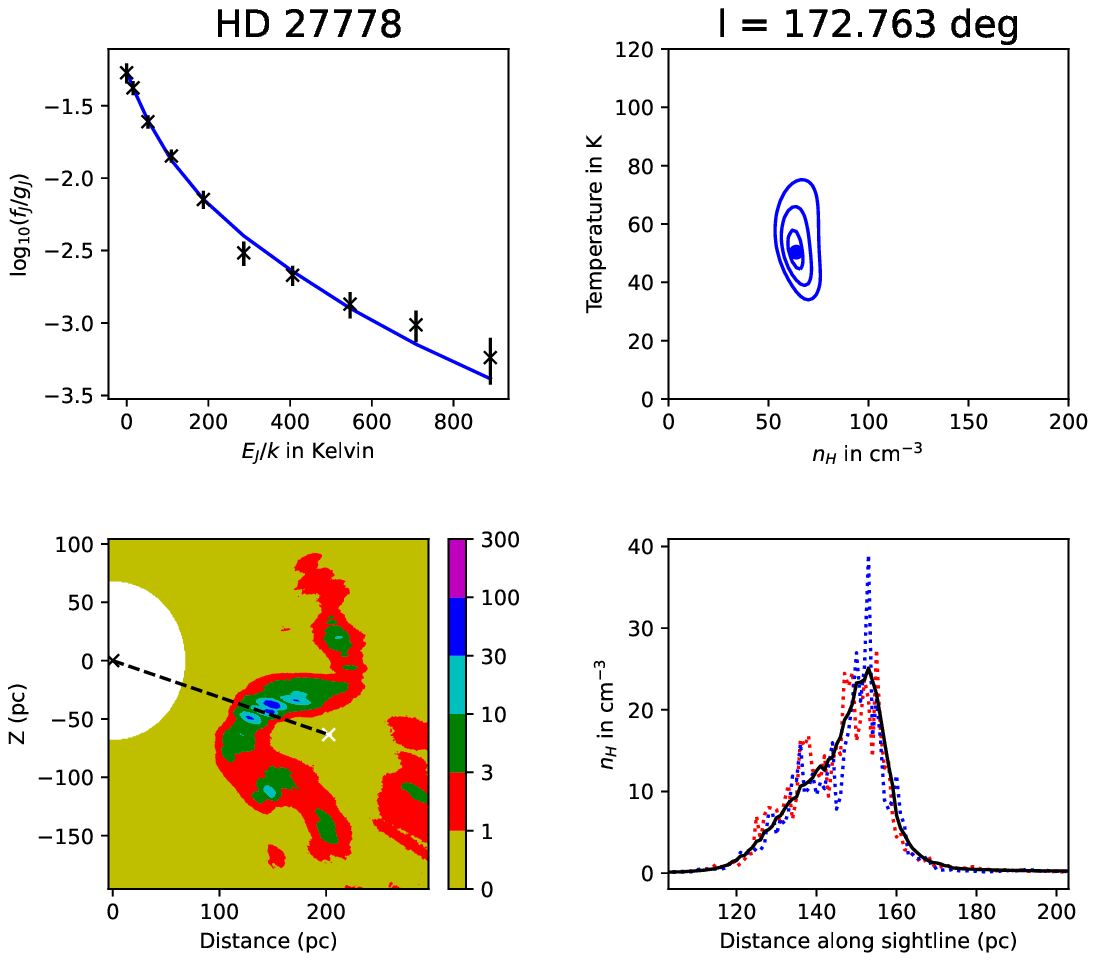}
\figurenum{10.2}
\figcaption{Results for the line-of-sight to HD 27778}
\vskip 4.1 true in
\end{figure}

\begin{figure}[!]
\includegraphics[angle=0,width=6 true in]{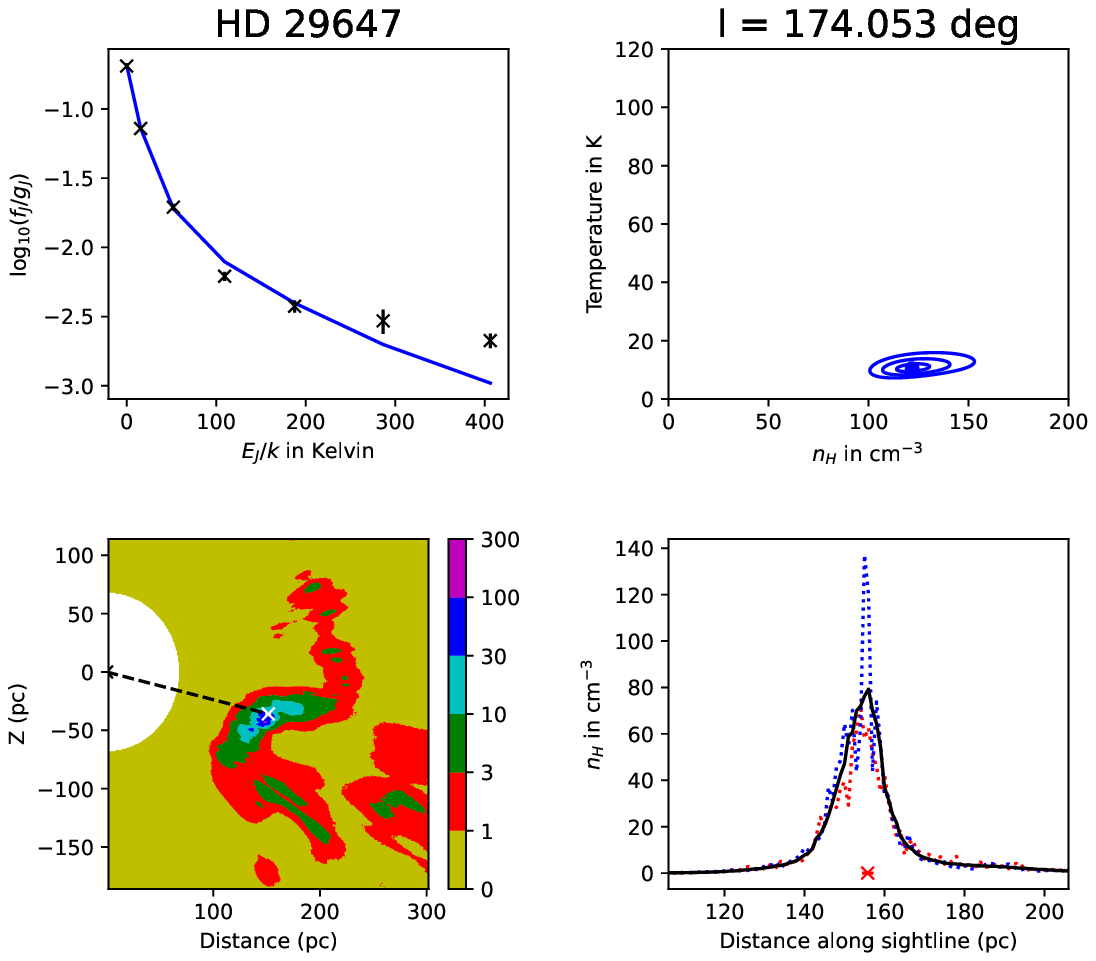}
\figurenum{10.3}
\figcaption{Results for the line-of-sight to HD 29647}
\vskip 4.1 true in
\end{figure}

\begin{figure}[!]
\includegraphics[angle=0,width=6 true in]{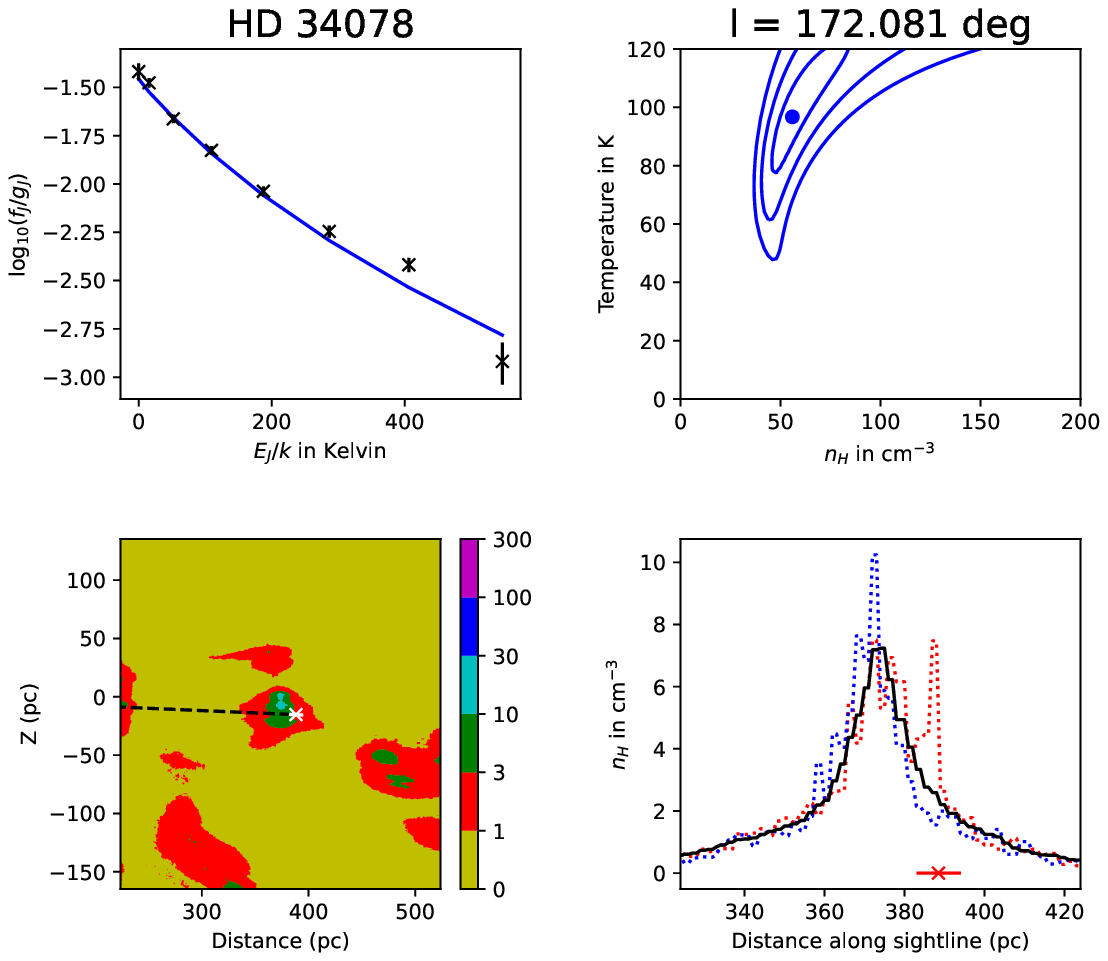}
\figurenum{10.4}
\figcaption{Results for the line-of-sight to HD 34078}
\vskip 4.1 true in
\end{figure}

\begin{figure}[!]
\includegraphics[angle=0,width=6 true in]{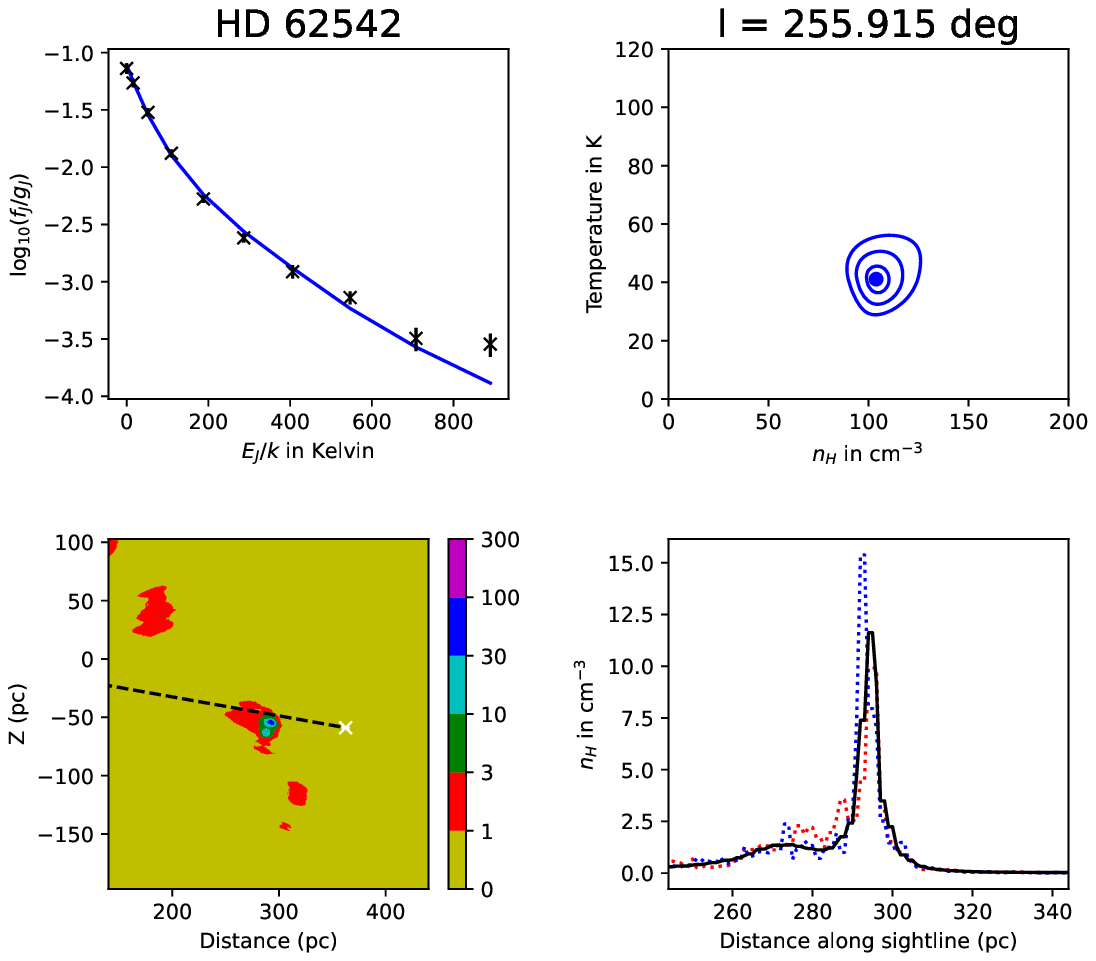}
\figurenum{10.5}
\figcaption{Results for the line-of-sight to HD 62542}
\vskip 4.1 true in
\end{figure}

\begin{figure}[!]
\includegraphics[angle=0,width=6 true in]{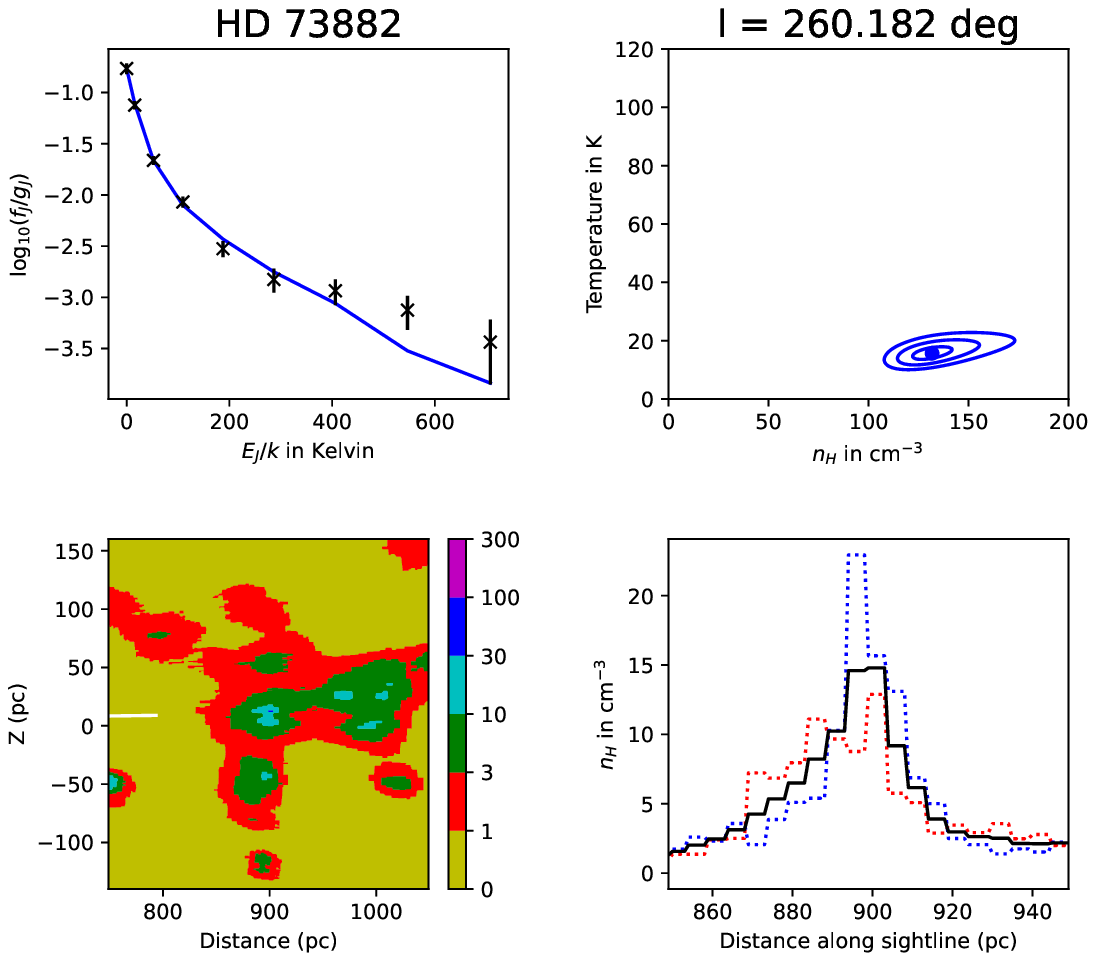}
\figurenum{10.6}
\figcaption{Results for the line-of-sight to HD 73882}
\vskip 4.1 true in
\end{figure}

\begin{figure}
\includegraphics[angle=0,width=6 true in]{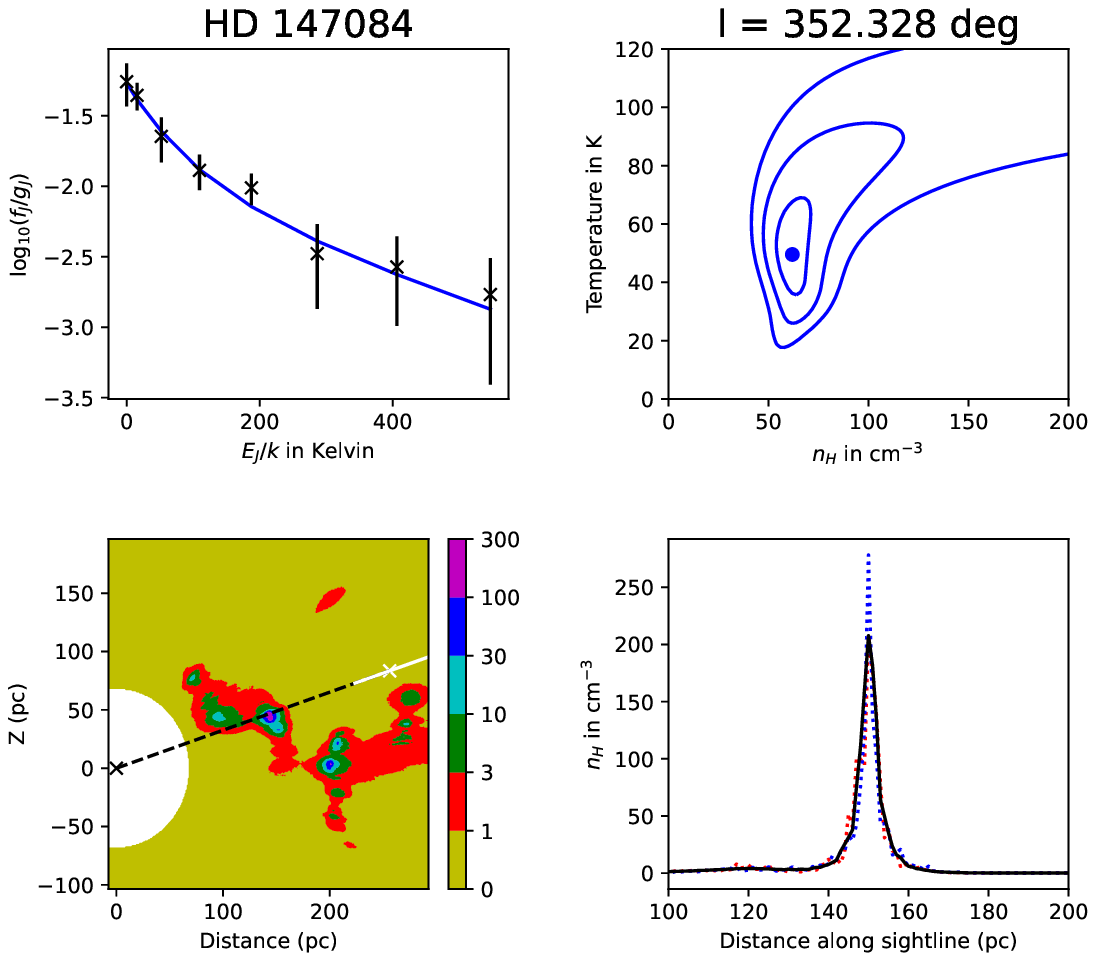}
\figurenum{10.7}
\figcaption{Results for the line-of-sight to HD 147084}
\vskip 4.1 true in
\end{figure}

\begin{figure}
\includegraphics[angle=0,width=6 true in]{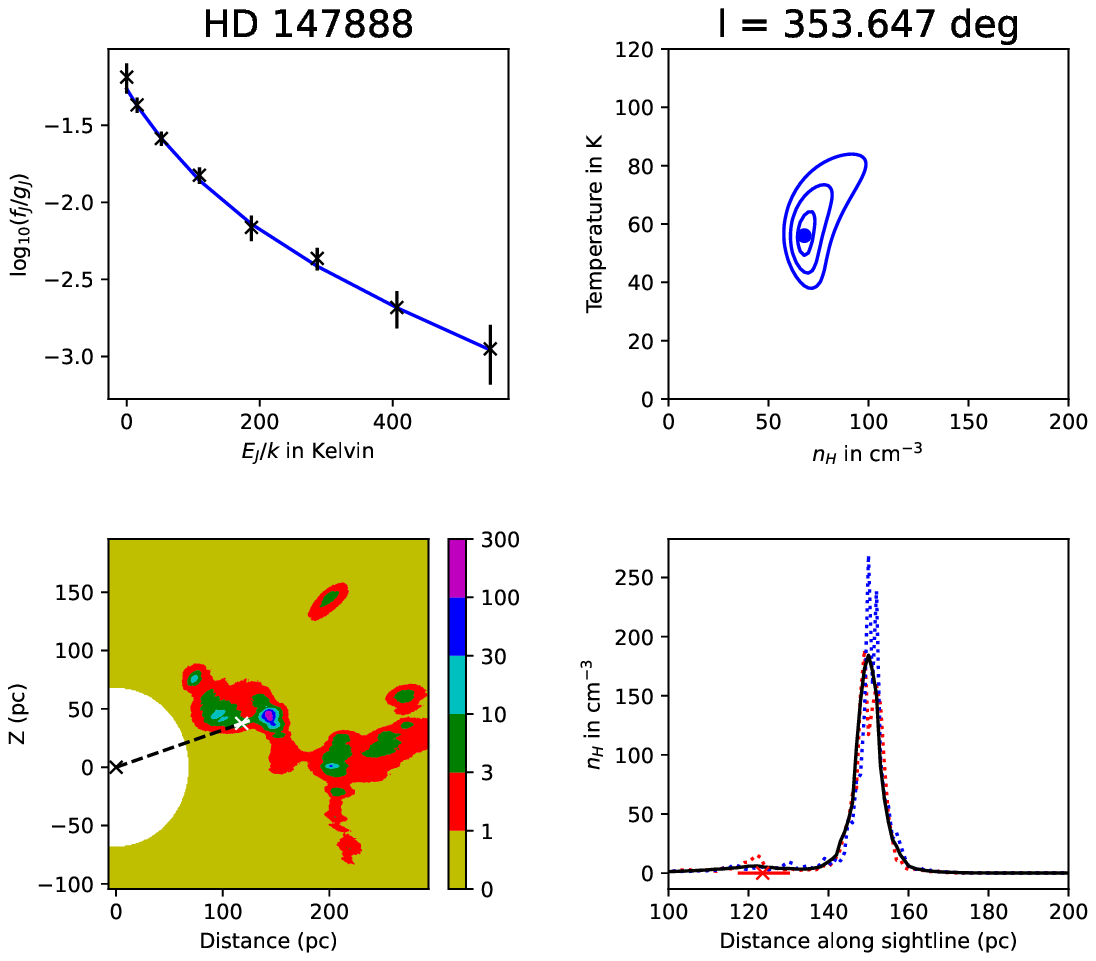}
\figurenum{10.8}
\figcaption{Results for the line-of-sight to HD 147888}
\vskip 4.1 true in
\end{figure}

\begin{figure}
\includegraphics[angle=0,width=6 true in]{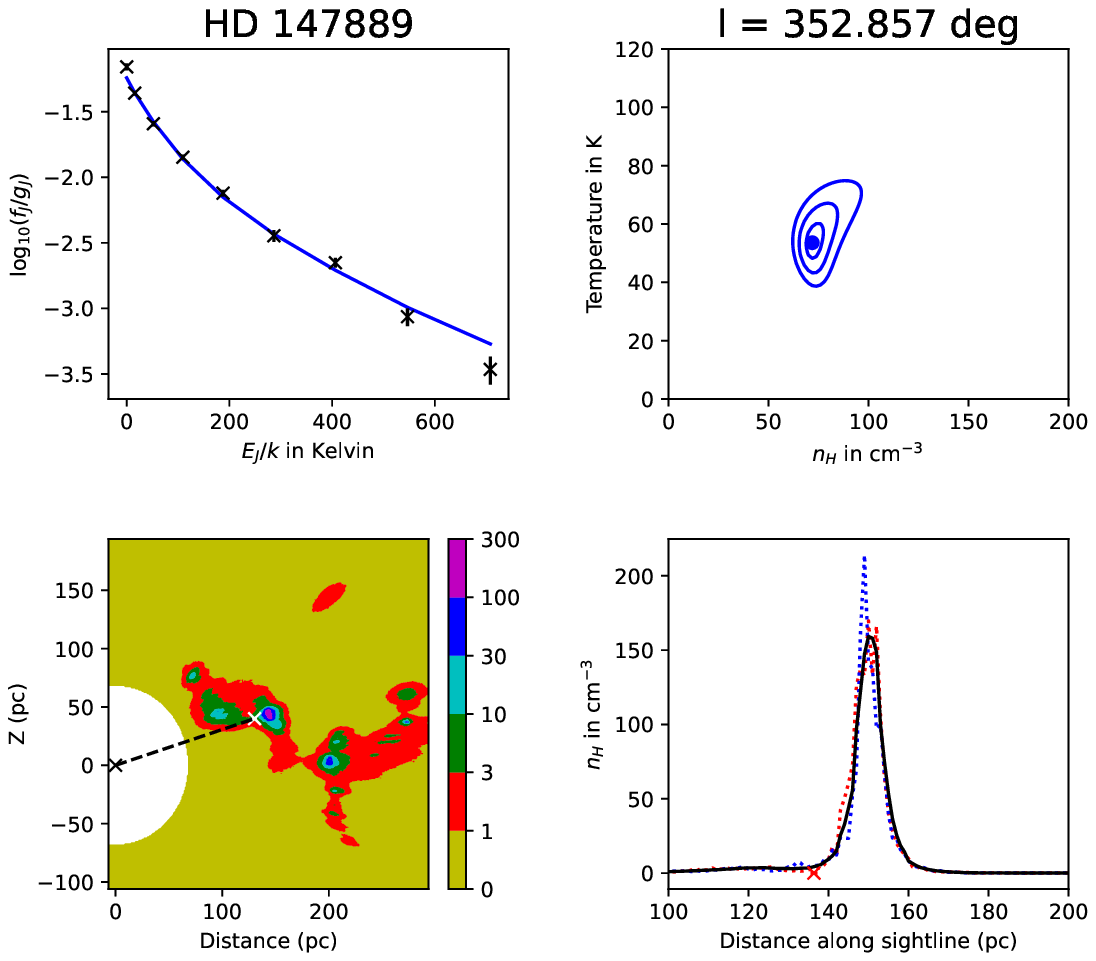}
\figurenum{10.9}
\figcaption{Results for the line-of-sight to HD 147889}
\vskip 4.1 true in
\end{figure}

\begin{figure}
\includegraphics[angle=0,width=6 true in]{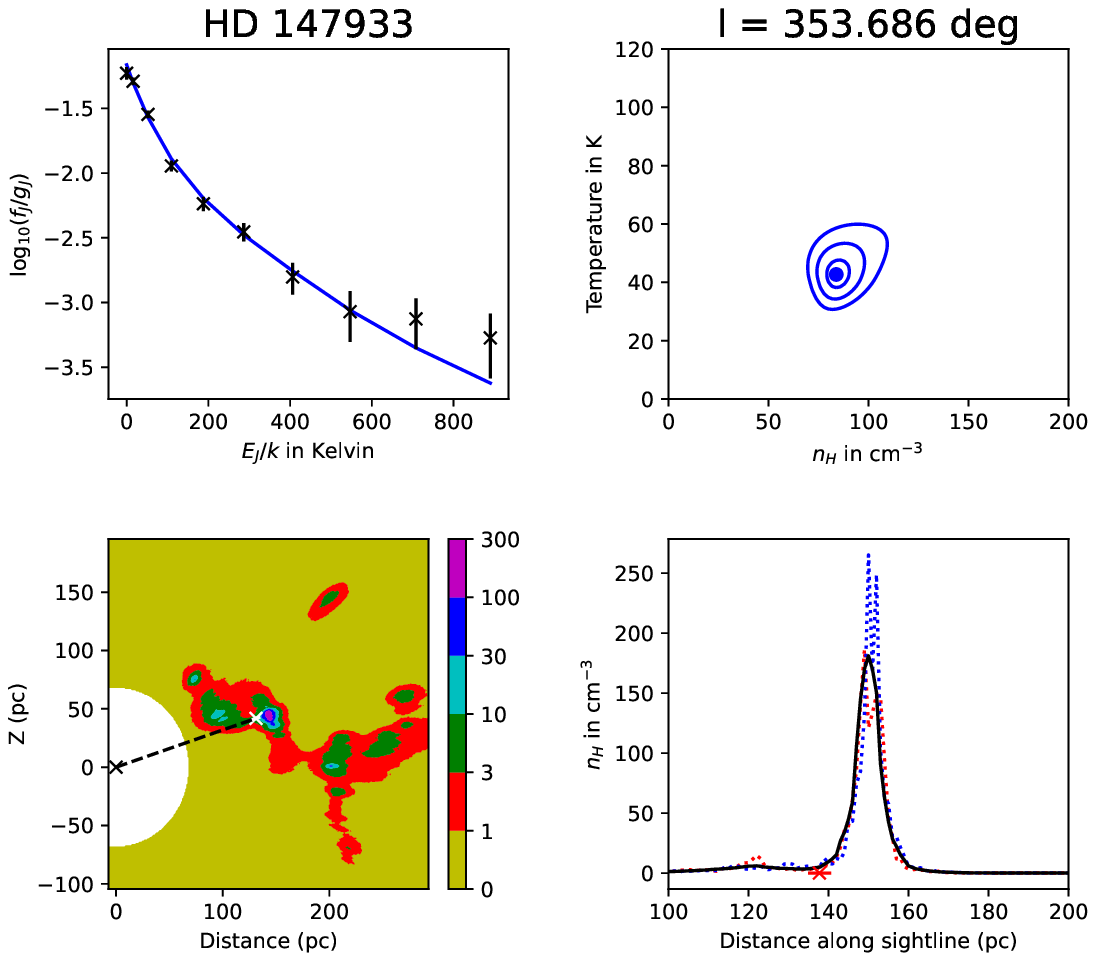}
\figurenum{10.10}
\figcaption{Results for the line-of-sight to HD 147933}
\vskip 4.1 true in
\end{figure}

\begin{figure}
\includegraphics[angle=0,width=6 true in]{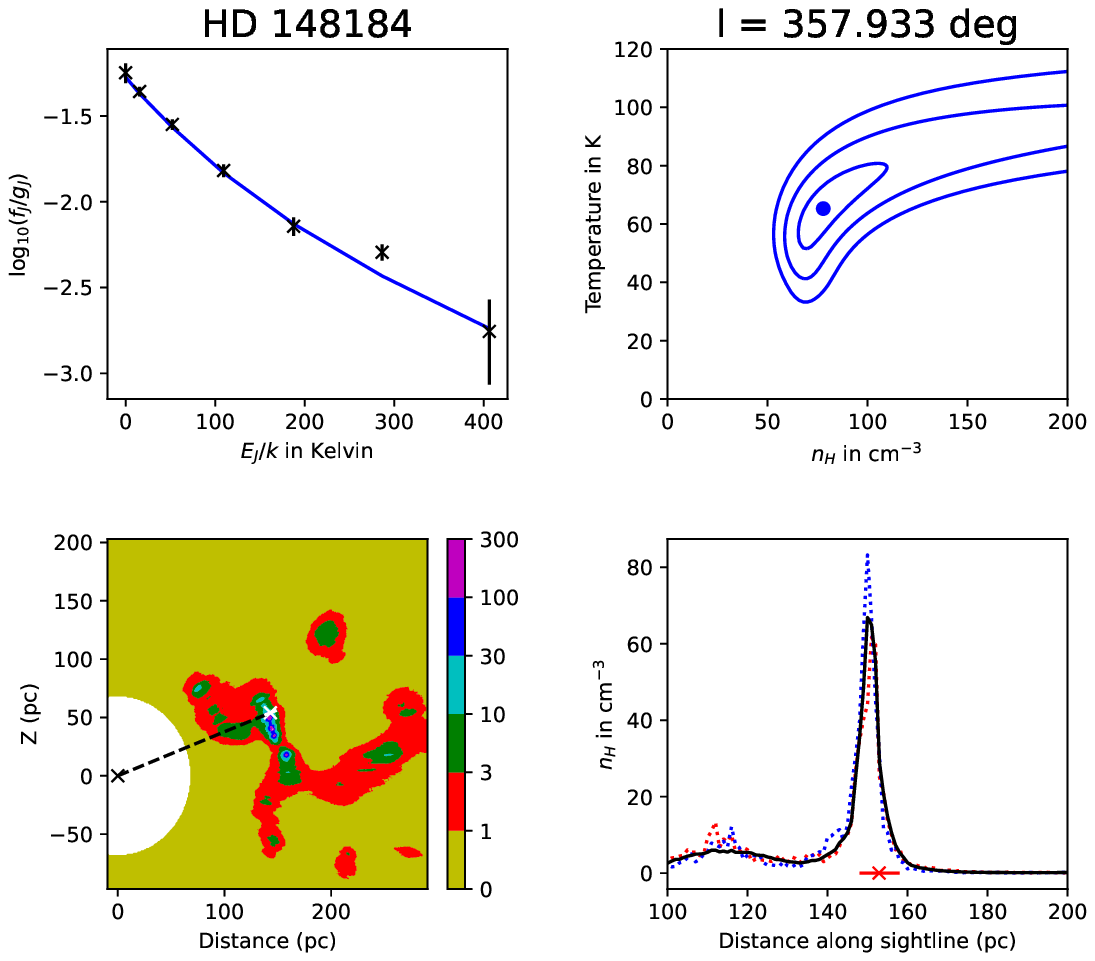}
\figurenum{10.11}
\figcaption{Results for the line-of-sight to HD 148184}
\vskip 4.1 true in
\end{figure}

\begin{figure}
\includegraphics[angle=0,width=6 true in]{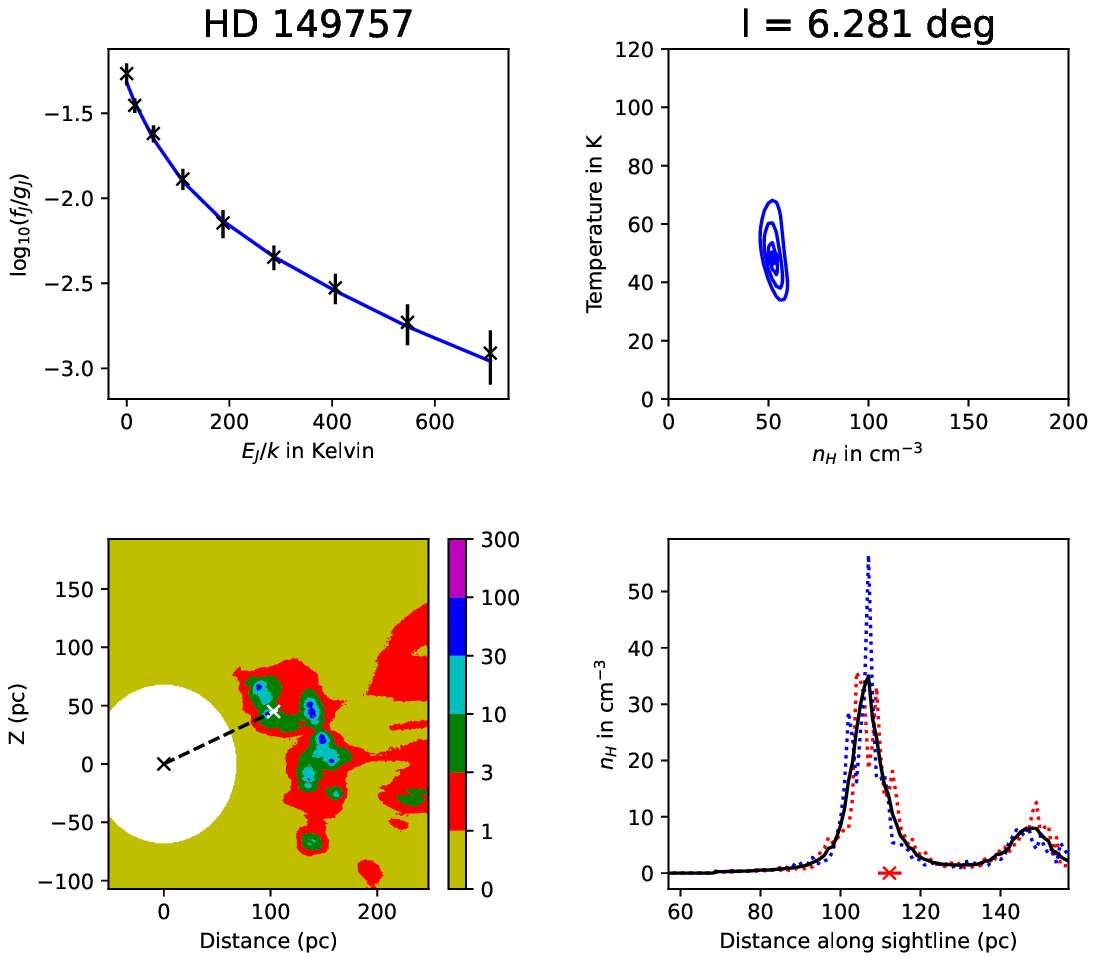}
\figurenum{10.12}
\figcaption{Results for the line-of-sight to HD 149757}
\vskip 4.1 true in
\end{figure}

\begin{figure}
\includegraphics[angle=0,width=6 true in]{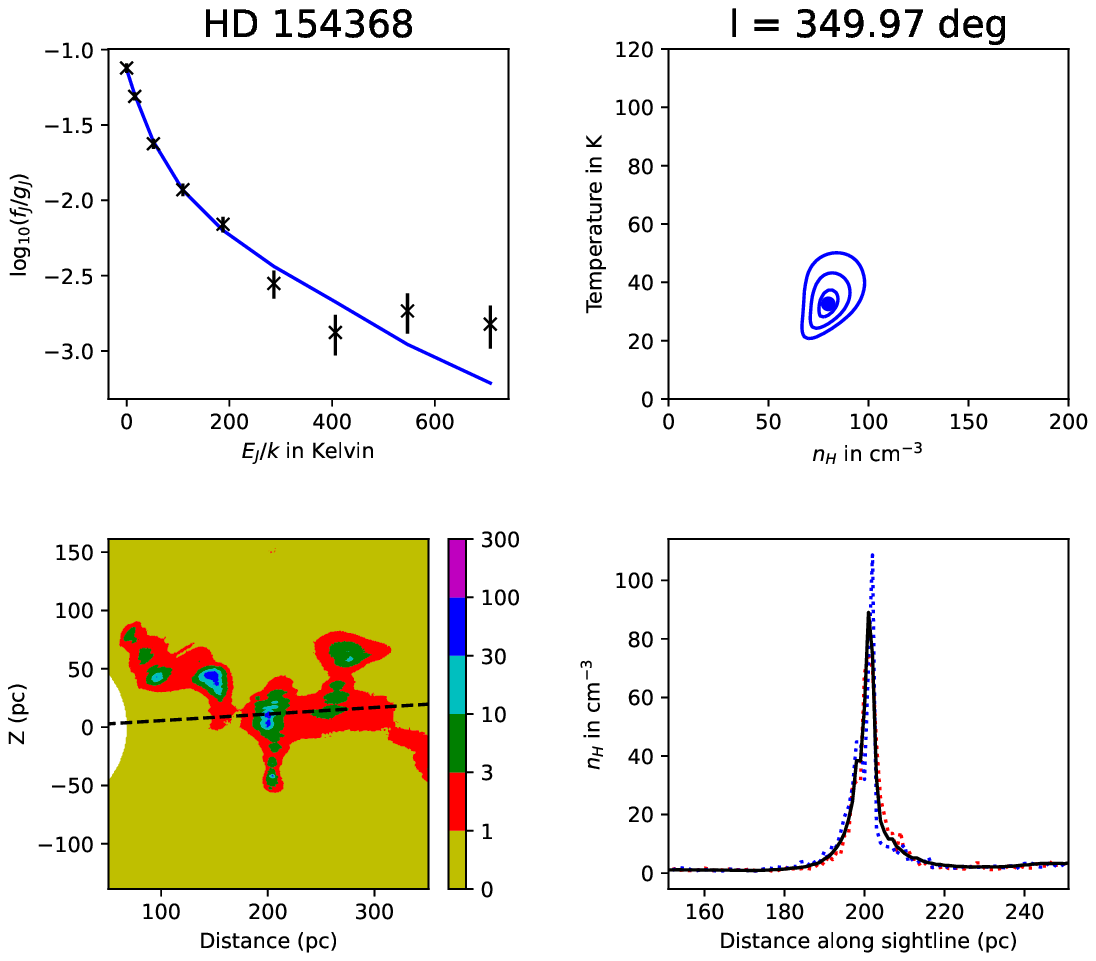}
\figurenum{10.13}
\figcaption{Results for the line-of-sight to HD 154368}
\vskip 4.1 true in
\end{figure}

\begin{figure}
\includegraphics[angle=0,width=6 true in]{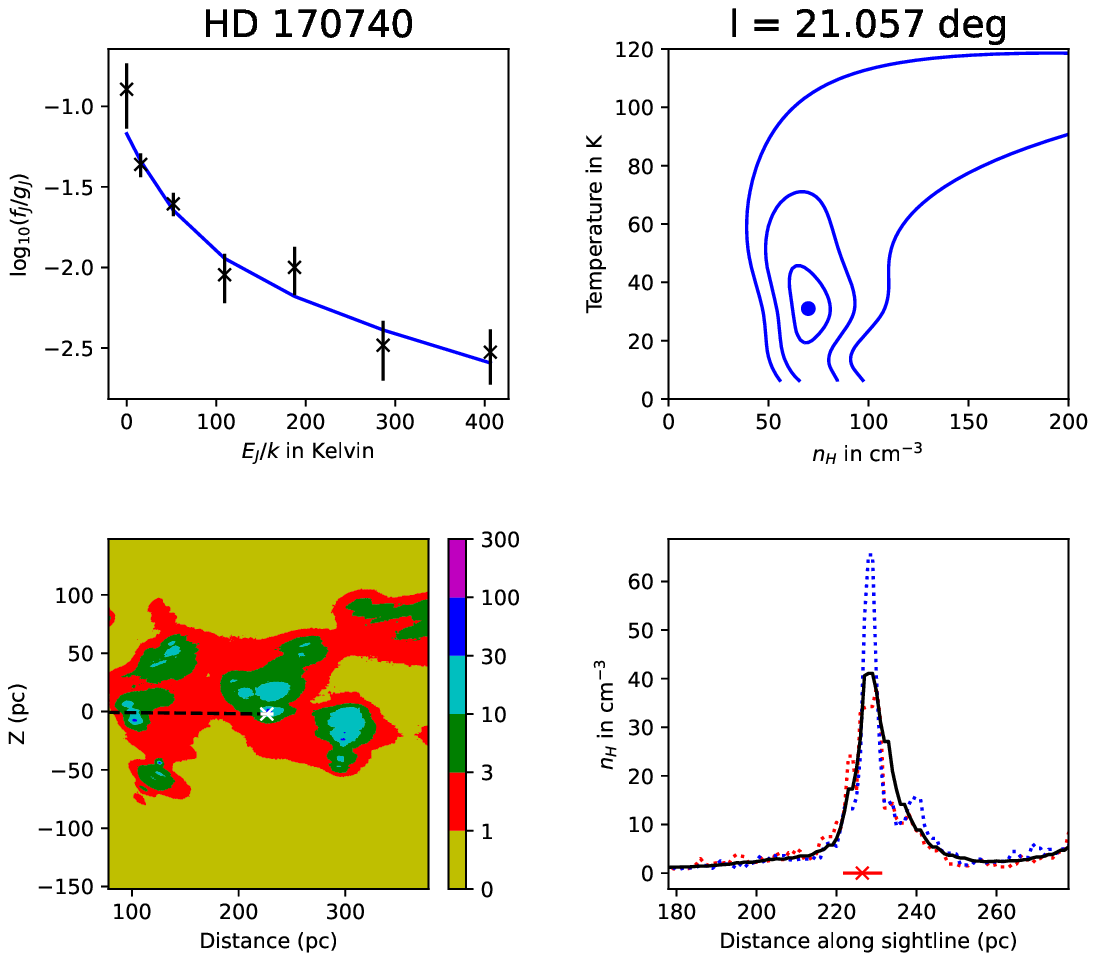}
\figurenum{10.14}
\figcaption{Results for the line-of-sight to HD 170740}
\vskip 4.1 true in
\end{figure}

\begin{figure}
\includegraphics[angle=0,width=6 true in]{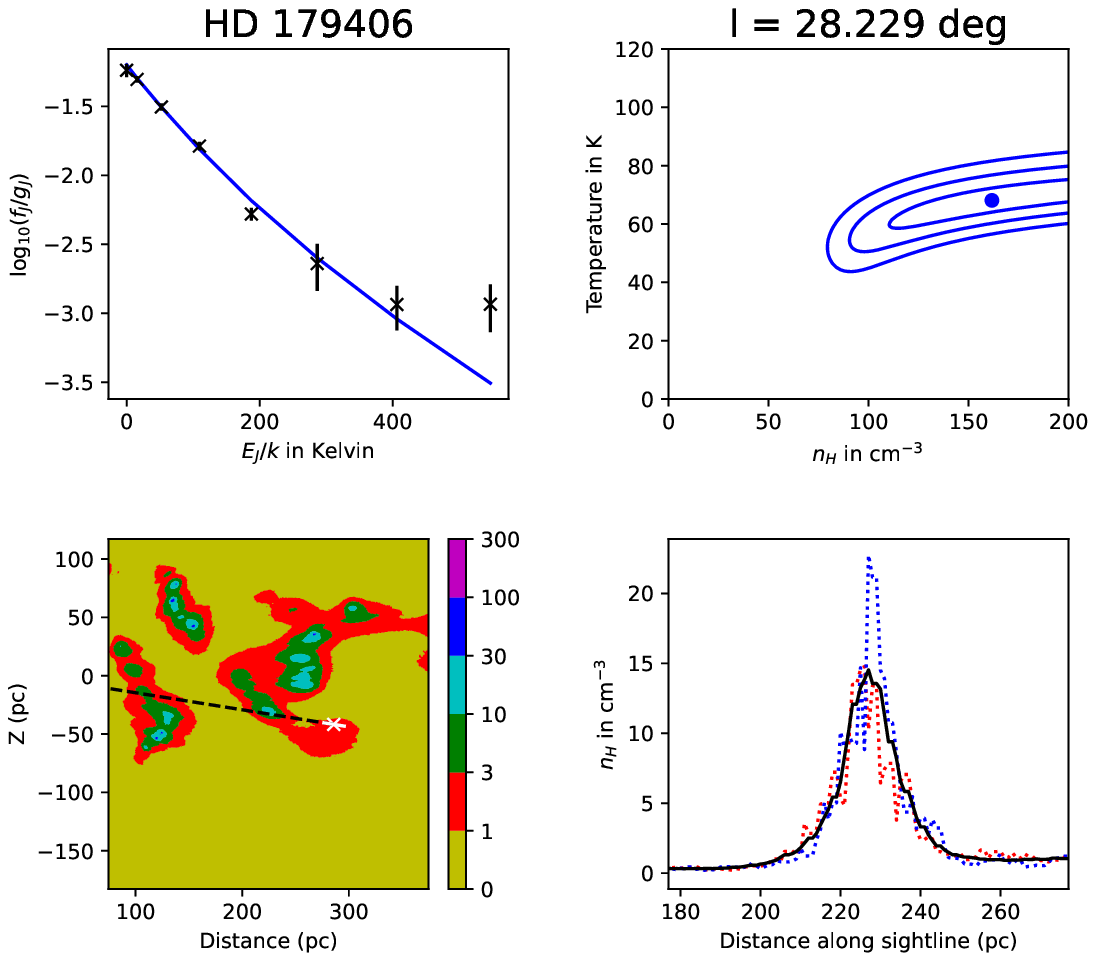}
\figurenum{10.15}
\figcaption{Results for the line-of-sight to HD 179406}
\vskip 4.1 true in
\end{figure}

\begin{figure}[b]
\includegraphics[angle=0,width=6 true in]{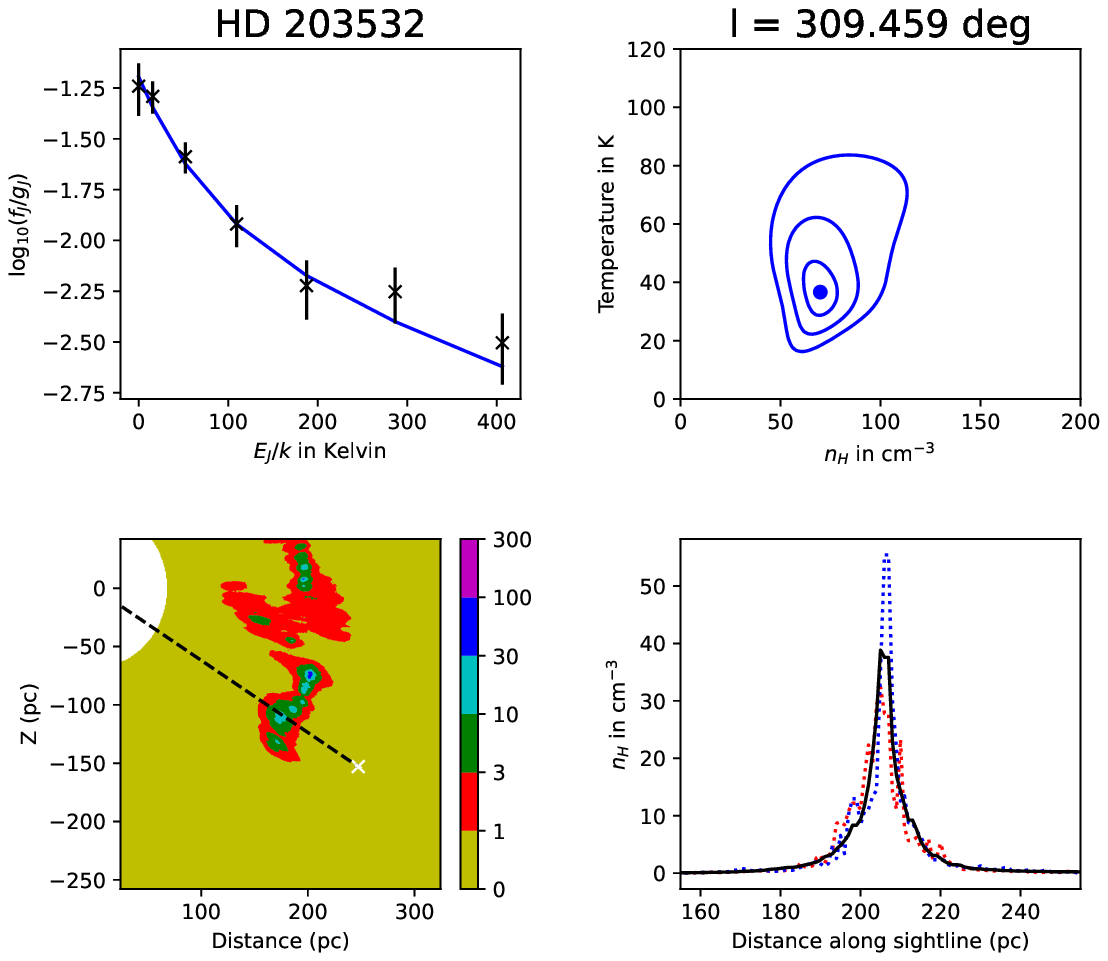}
\figurenum{10.16}
\figcaption{Results for the line-of-sight to HD 203532}
\vskip 4.1 true in
\end{figure}

\begin{figure}
\includegraphics[angle=0,width=6 true in]{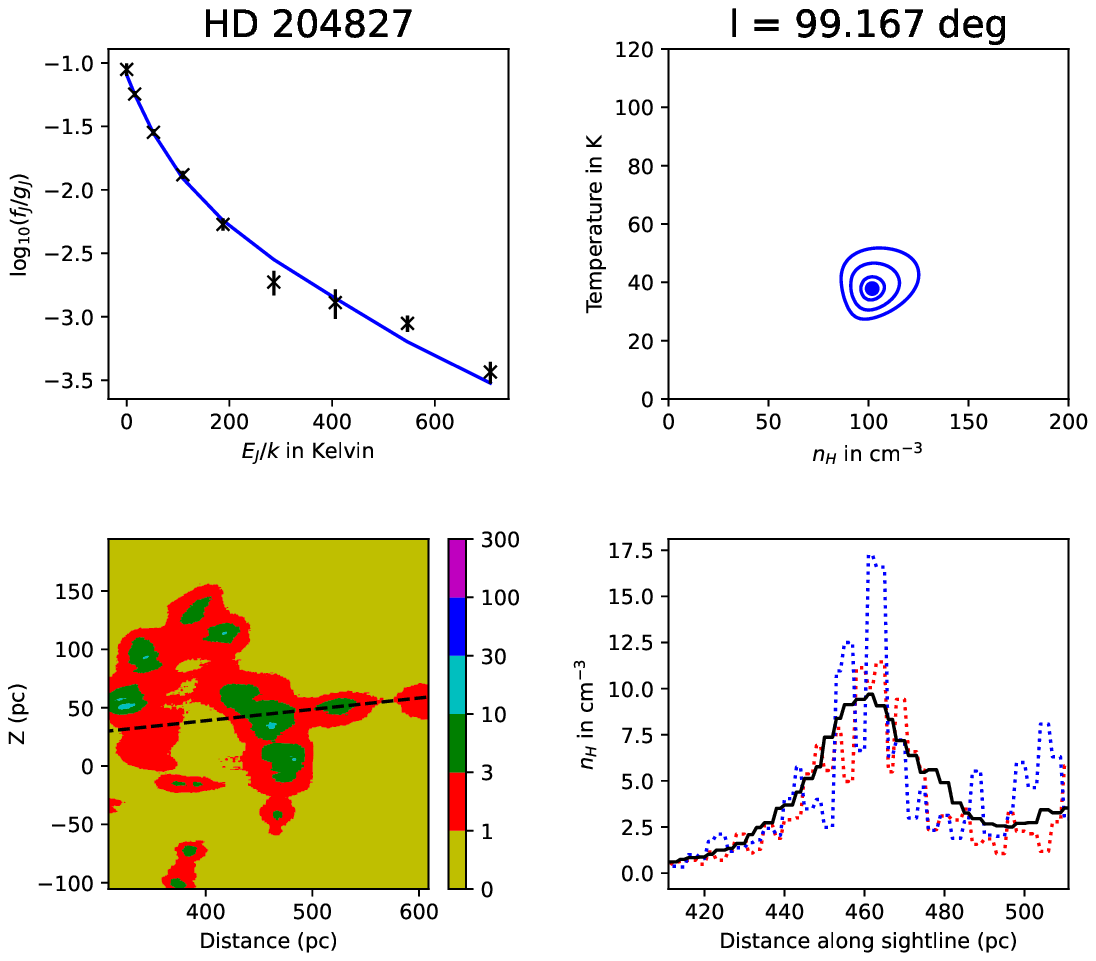}
\figurenum{10.17}
\figcaption{Results for the line-of-sight to HD 204827}
\vskip 4.1 true in
\end{figure}

\begin{figure}
\includegraphics[angle=0,width=6 true in]{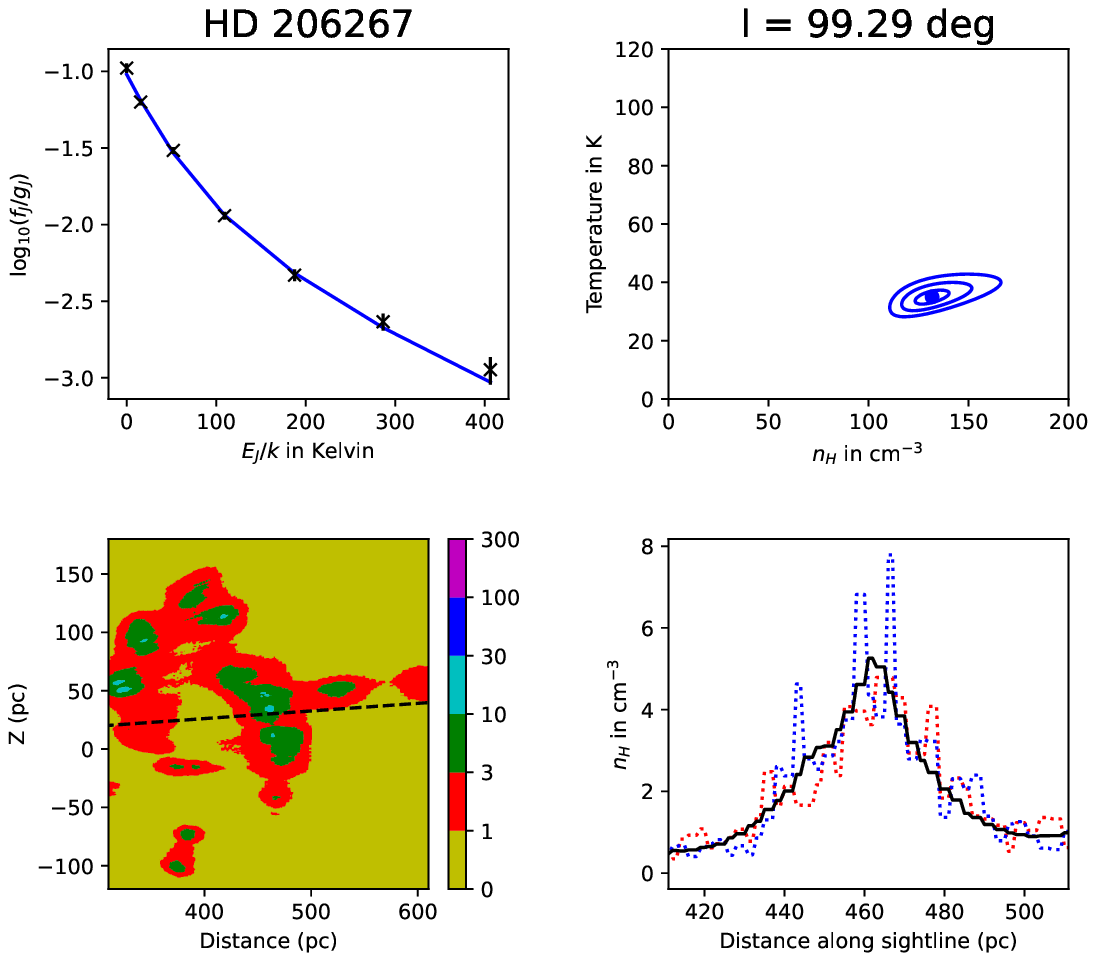}
\figurenum{10.18}
\figcaption{Results for the line-of-sight to HD 206267}
\vskip 4.1 true in
\end{figure}

\begin{figure}
\includegraphics[angle=0,width=6 true in]{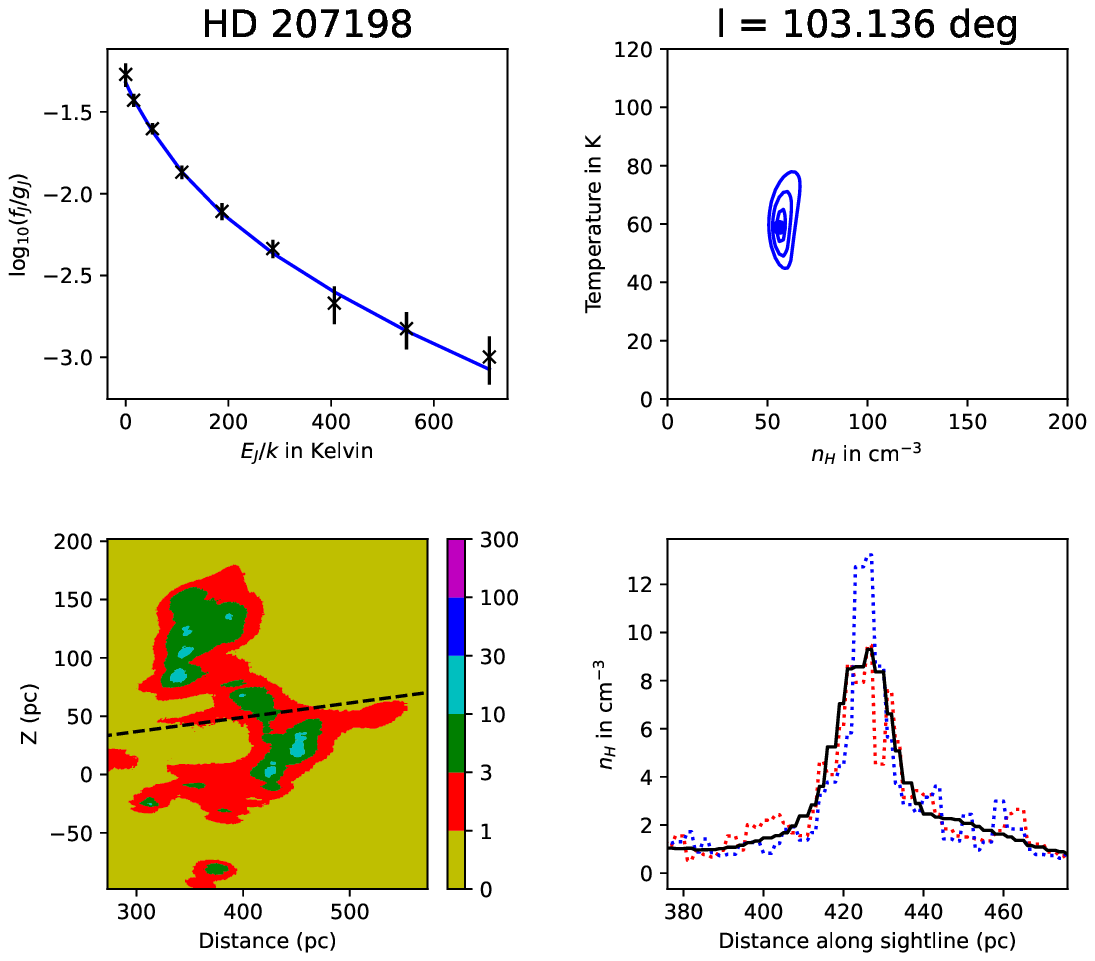}
\figurenum{10.19}
\figcaption{Results for the line-of-sight to HD 207198}
\vskip 4.1 true in
\end{figure}

\begin{figure}
\includegraphics[angle=0,width=6 true in]{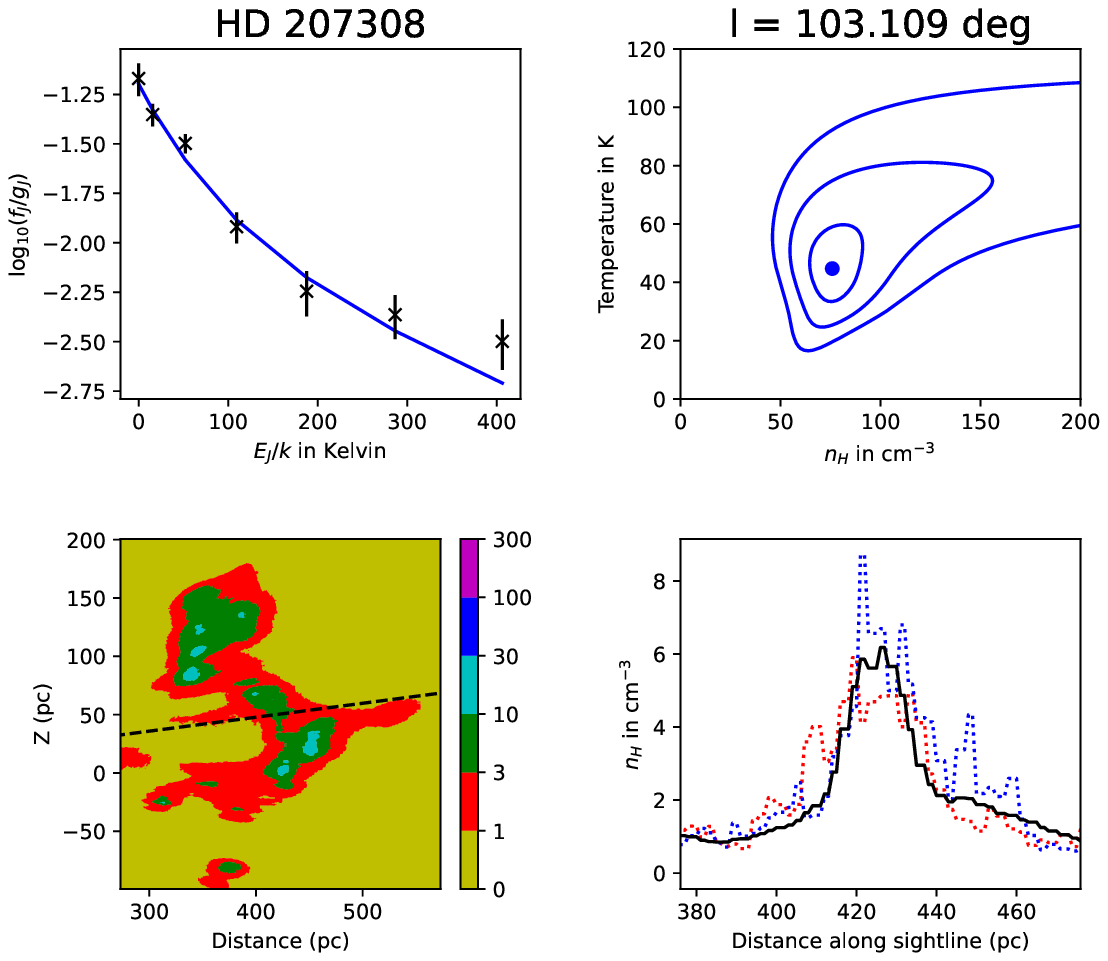}
\figurenum{10.20}
\figcaption{Results for the line-of-sight to HD 207308}
\vskip 4.1 true in
\end{figure}

\begin{figure}
\includegraphics[angle=0,width=6 true in]{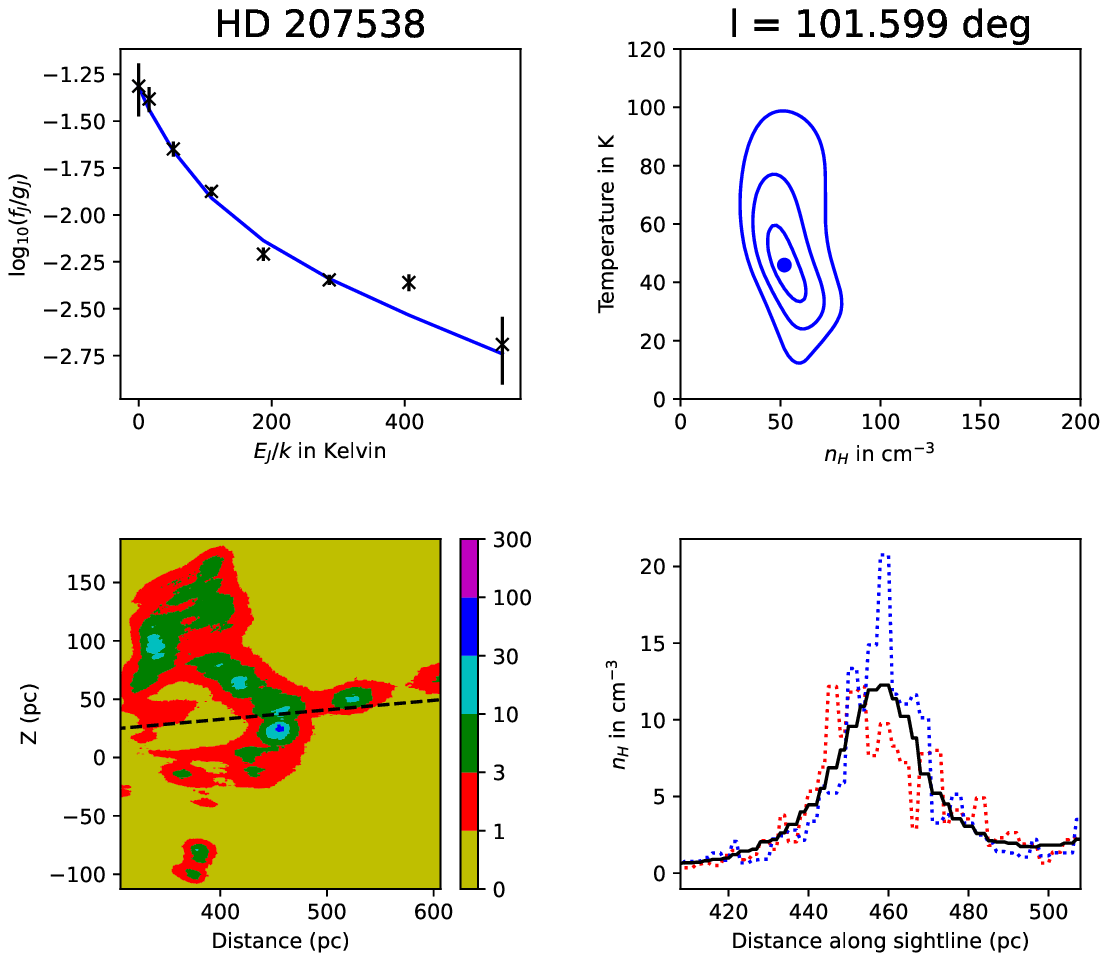}
\figurenum{10.21}
\figcaption{Results for the line-of-sight to HD 207538}
\vskip 4.1 true in
\end{figure}

\begin{figure}
\includegraphics[angle=0,width=6 true in]{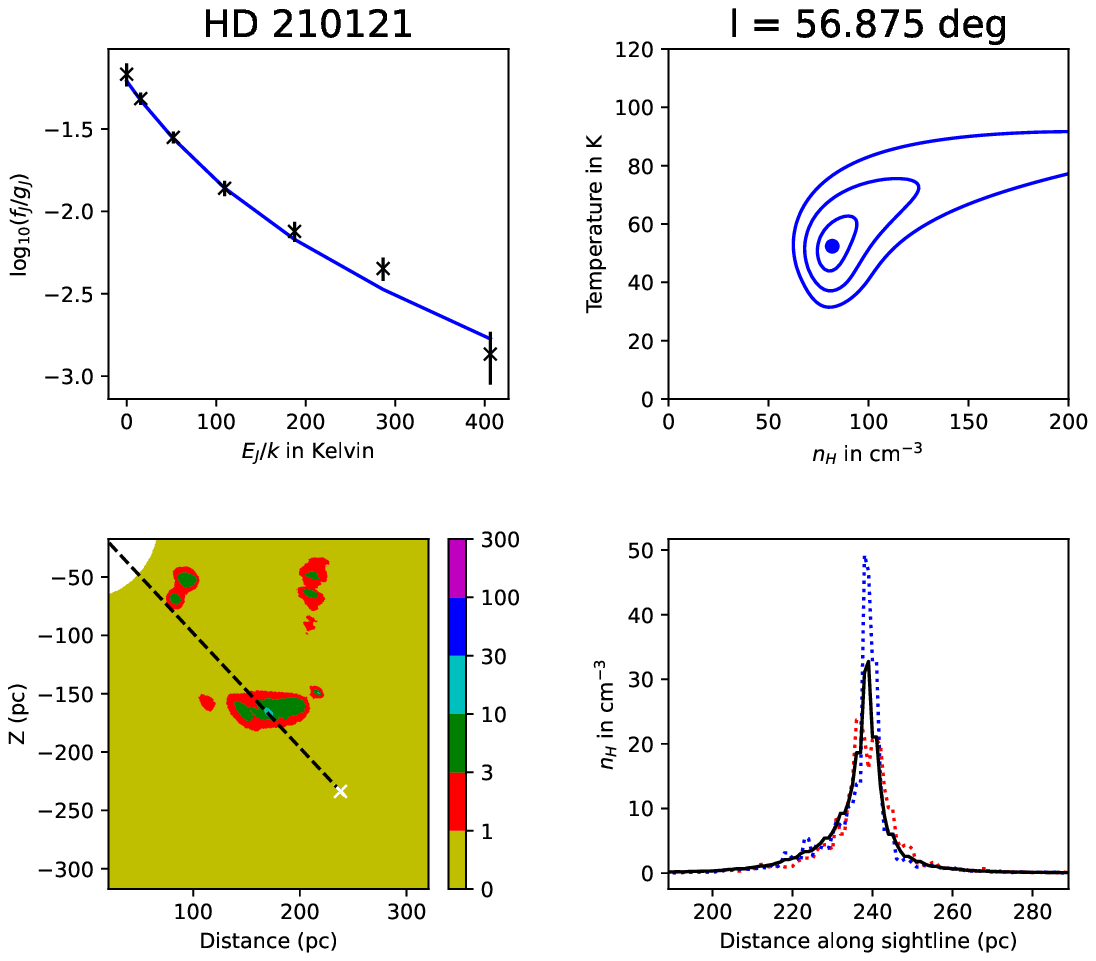}
\figurenum{10.22}
\figcaption{Results for the line-of-sight to HD 210121}
\vskip 4.1 true in
\end{figure}

\begin{figure}
\includegraphics[angle=0,width=6 true in]{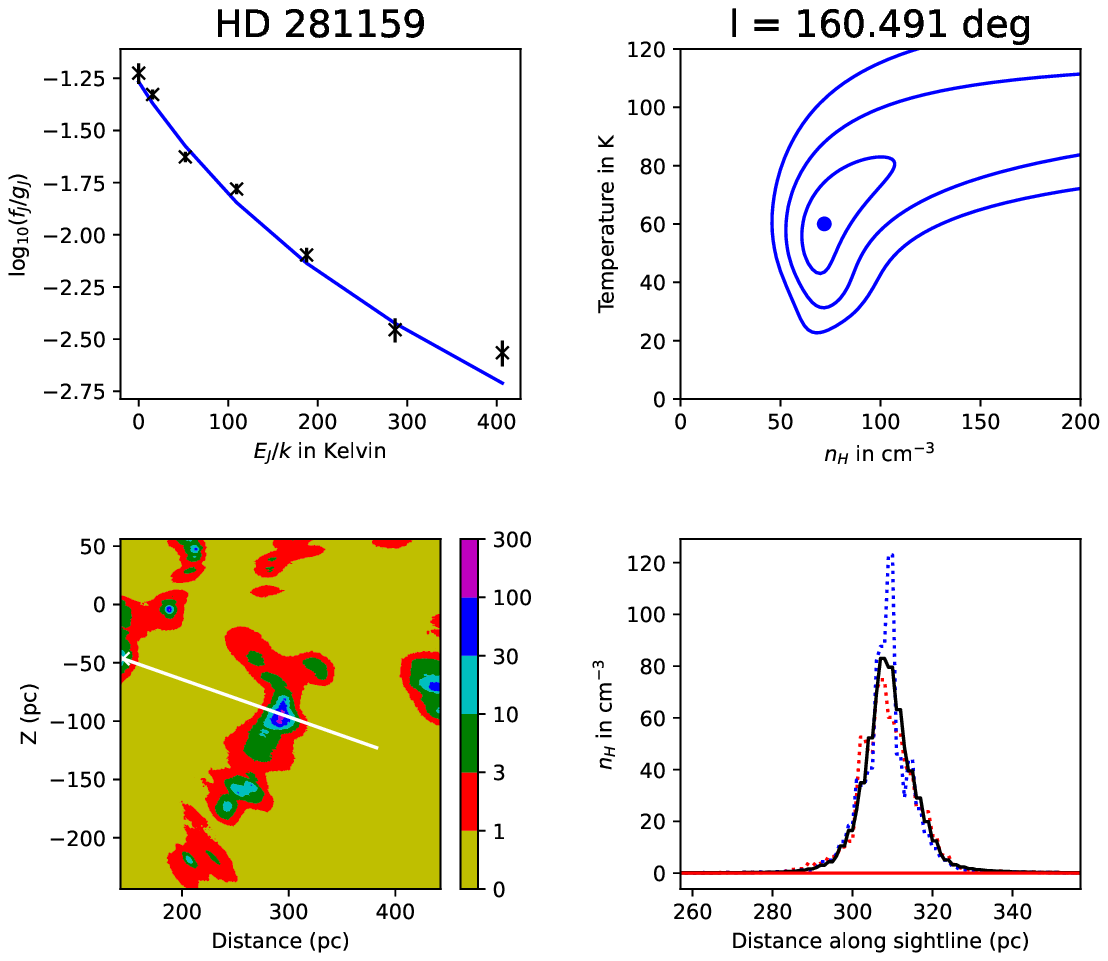}
\figurenum{10.23}
\figcaption{Results for the line-of-sight to HD 281159}
\vskip 4.1 true in
\end{figure}

\end{document}